\documentclass[article,showpacs,amsmath,amssymb,nofootinbib]{revtex4}
\usepackage[utf8]{inputenc}
\usepackage[english]{babel}
\usepackage{graphicx}
\usepackage{empheq}

\newcommand{\om} \omega   
\newcommand{\Om} \Omega
\newcommand{\eps} \epsilon
\newcommand{\la} \lambda

\newcommand{\tal} {\tilde\alpha}
\newcommand{\tga} {\tilde \gamma}

\newcommand{\tT}{\tilde T}
\newcommand{\tR}{\tilde R}

\newcommand{\be} {\begin{equation}}
\newcommand{\ee} {\end{equation}}

 \newcommand{\ba} {\begin{eqnarray}}
 \newcommand{\ea} {\end{eqnarray}}

\newcommand{\bal} {\begin{align}}
\newcommand{\eal} {\end{align}}

\newcommand{\sba}{\begin{subeqnarray}}
\newcommand{\sea}{\end{subeqnarray}}

\def\lrD{\mathrel{{\cal D}\kern-1.em\raise1.75ex\hbox{$\leftrightarrow$}}}
\def\lr #1{\mathrel{#1\kern-1.25em\raise1.75ex\hbox{$\leftrightarrow$}}}
\newcommand{\eqr}[1]{Eq.~(\ref{#1})}

\begin{document}
\title{Black hole lasers, a mode analysis.}

\author{Antonin Coutant}
\email{antonin.coutant@th.u-psud.fr}
\affiliation{Laboratoire de Physique Th\'eorique, CNRS UMR 8627, B\^at. 210, Universit\'e Paris-Sud 11, 91405 Orsay Cedex, France}
\author{Renaud Parentani}
\email{renaud.parentani@th.u-psud.fr}
\affiliation{Laboratoire de Physique Th\'eorique, CNRS UMR 8627, B\^at. 210, Universit\'e Paris-Sud 11, 91405 Orsay Cedex, France}

\date{\today}

\begin{abstract}
We show that the black hole laser effect discovered by Corley $\&$ Jacobson should be described in terms of 
frequency eigenmodes that are spatially bound. The spectrum contains 
a discrete and finite set 
of complex frequency modes which appear in pairs 
and which encode 
the laser effect.
In addition, it contains 
real frequency modes that form a continuous set 
when space is infinite, and which 
are only elastically scattered, 
i.e., not subject to any Bogoliubov transformation.
 The quantization is straightforward, 
but the calculation of the 
asymptotic fluxes is rather involved. 
When the number of complex frequency modes is small, our expressions
differ from those given earlier. 
In particular, when the region between the horizons shrinks, 
there is a minimal distance 
under which no complex frequency mode exists, and no radiation is emitted.
Finally, we relate 
this effectto other dynamical instabilities found for rotating black holes and 
in electric fields, 
and we give the conditions to get this type of instability.

\end{abstract}



\maketitle

\section{Introduction}

In 1998, Corley and Jacobson 
made the following interesting observation~\cite{Corley:1998rk}.
They noticed that the propagation of a superluminal dispersive field in a stationary geometry
containing two horizons, as it is the case for nonextreme charged black holes,
leads to a self amplified Hawking radiation (for bosonic fields). The origin
of this laser effect can be attributed 
to the closed 
trajectories followed
by the negative Killing frequency partners of Hawking quanta. Because of the superluminal
character of the dispersion, the partners indeed bounce from one horizon to the other.

Besides charged black holes, 
the acoustic geometries associated with a moving fluid that crosses twice the speed of sound
also contain 
a pair of horizons.
For one-dimensional flows these 
geometries are 
of the form
\be
ds^2 = -dt^2 + (dx - v(x)dt)^2.\label{metric}
\ee
This expression obtains~\cite{Unruh:1980cg,Barcelo:2005fc} when considering the 
propagation of low frequency phonons in a moving fluid of 
velocity $v(x)$.
The sound velocity is assumed to be constant and has been set to $1$. 
This metric possesses a black hole (BH) or a white hole (WH) horizon when the gradient $\partial_x v$ 
at the horizon $\vert v \vert = 1$
is positive or negative.
When the 
dispersion is superluminal (anomalous) all such systems should experience a lasing effect. 
This should apply 
in particular 
to the BH-WH pair realized 
in a Bose Einstein condensate~\cite{Jeff}.

The original analysis and that of 
\cite{Leonhardt:2008js}
were  both 
carried out using wave packets. 
In the present work we show
that there exists a 
more fundamental 
description 
based on frequency eigenmodes which are asymptotically bound (in space).
When 
$v(x)$ is 
constant and subsonic for both $x\to \pm  \infty$, 
there is a discrete set of complex frequency modes
and a continuous family of real frequency modes.
(If instead the subsonic region 
is finite
and periodic conditions imposed, 
the situation is more complicated~\cite{Gardiner,Garay:2000jj} and will not be considered here.)
In our case, the real frequency modes are asymptotically oscillating and normalized by a delta of Dirac. 
They are not subject to any Bogoliubov transformation. In fact, 
the scattering matrix at fixed $\om$ 
only
contains reflection and transmission coefficients mixing right and left moving (positive norm) modes.
This was not {\it a priori} expected since
the 
matrix associated with a single BH (or a 
WH)
is 
$3 \times 3$ 
and mixes positive and negative norm modes in a nontrivial way~\cite{Macher:2009tw,Macher:2009nz}.

The discrete set is composed of modes that vanish for $x \to \pm \infty$. 
They
 form two modes subsets of complex conjugated frequencies, 
each containing a growing and a decaying mode. 
The time dependence of the coefficients in each subset corresponds to that 
of a complex upside down and rotating harmonic oscillator. 
Both the real and the imaginary part of the frequency play important
roles in determining the asymptotic properties of the fluxes. 
We notice such modes were encountered
in several other situations~\cite{Fulling:1989nb,Damour:1976kh,Damour:1978ug,Kang:1997uw,Cardoso:2004nk,Greiner}.
We also mention that a stability analysis of BH-WH flows in Bose Einstein condensates
was presented in \cite{Barcelo:2006yi}. 
We reach different conclusions
because we use different boundary conditions.

In Sec.~\ref{settings}, we present our settings.
In Sec.~\ref{setofmodes}, we demonstrate that the 
set of 
spatially bound modes contains a continuous part and a discrete part 
composed of complex frequency modes. 
In Sec.~\ref{modeprop} and Sec.~\ref{predic2}, 
we study the properties of the modes,
and show how the 
complex frequency modes determines the fluxes.
We also relate our approach to that
based on wave packets~\cite{Corley:1998rk,Leonhardt:2008js},
and explain why the predictions differ in general, and in particular 
when the number of discrete 
modes is small. In Sec.~\ref{condi} we give the conditions to get 
complex frequency modes in general terms. 


\section{The settings\label{settings}}

\begin{figure} 
\includegraphics[
height=80mm]{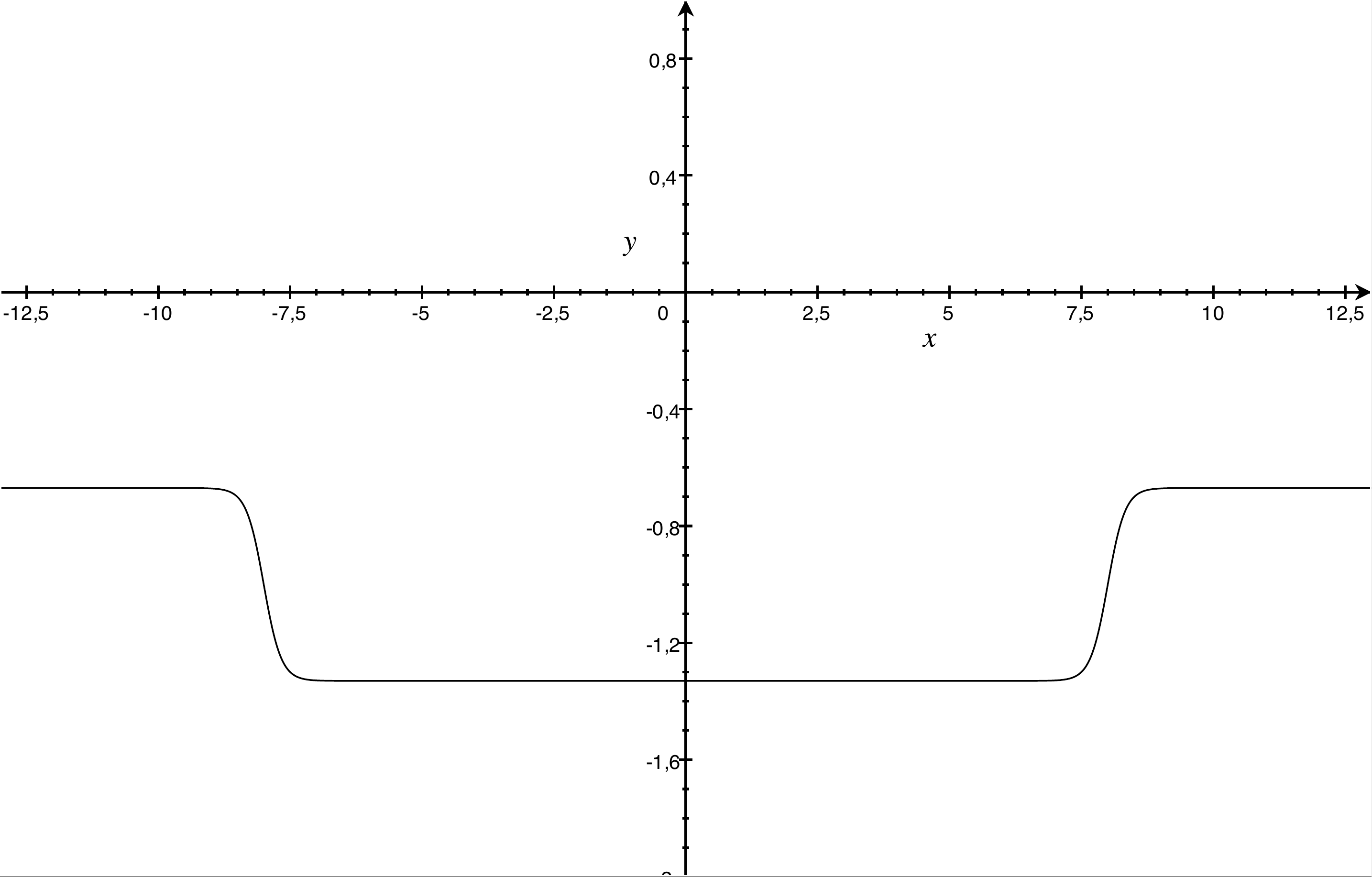}
\caption{Velocity profile $v(x)$ 
as a function of $\kappa x$, for $\kappa_B = \kappa_W$,
$D=0.33$, and $\kappa L=8$. 
The horizons are located at $\kappa x= \pm 8$, where $v(x) + 1 = 0$.}
\label{fig} 
\end{figure}

We work in two space-time dimensions and consider the stationary
metrics of \eqr{metric} which contain a BH-WH pair. 
We restrict ourselves to flows that are asymptotically constant, i.e.
we consider 
velocity profiles such as  
\be
v(x) = -1 + D \,
 \tanh\left[\frac{\kappa_B (x - L) }{D}\right] \,  
 \tanh\left[\frac{\kappa_W (x + L) }{D}\right] ,
\label{vdparam}
\ee
see Fig.~\ref{fig}.
The BH (WH) horizon is situated at $x=L$ ($x=-L$).
We 
suppose that the inequality
$\kappa L/D \gg 1$ is satisfied for both values of $\kappa$, where $D\in \, ]0,1]$.
In this case, the two near horizon regions of width $\Delta x \sim D/\kappa$
are well separated, 
and the surface gravities of the BH and
the WH are, respectively, given by
$\kappa_B= \partial_x v\vert_{x=L}$ and $\kappa_W = -\partial_x v\vert_{x=-L}$.
%
The minimal speed $\vert v_-\vert < 1$ is reached for $x \to \pm \infty$, whereas 
the maximal speed $\vert v_+ \vert > 1$ 
is found at $x=0$ between the two horizons.
When $\kappa L/D \gg 1$, their values 
 are
\be v_\pm = -1 \mp D.
\label{v+}
\ee
As emphasized in \cite{Macher:2009tw,Macher:2009nz}, when using nonlinear dispersion relations,
$D$ fixes the critical frequency $\om_{\rm max}$
above which no radiation is emitted by a single BH, or WH. 
Similarly here, there will be no unstable mode above $\om_{\rm max}$.

As in \cite{Corley:1998rk} 
we work with a real field $\phi$ obeying a quartic superluminal dispersion relation
\be
\Om^2 = k^2 + k^4/\Lambda^{2}. 
\label{disprela}
\ee
The UV scale characterizing the 
dispersion is given by $\Lambda$, 
and $\Om$ is the frequency measured in the preferred frame, that comoving with the fluid.
Most of the results we shall derive also apply to higher order superluminal dispersion relations.
They also apply to the Bogoliubov theory of phonons in Bose condensates 
which is a variant of the present case,
as can be verified by comparing \cite{Macher:2009nz} to \cite{Macher:2009tw}.
The action of $\phi$ in the metric of \eqr{metric} is
\ba
S = \frac{1}{2} \int dt dx \left[ \left(\partial_t \phi + v\partial_x \phi\right)^2 - (\partial_x \phi)^2
- \frac{1}{\Lambda^2}(\partial_x^2 \phi)^2 \right], 
\label{action}
\ea
and the wave equation 
is 
\ba
\left[(\partial_t+ \partial_x v)(\partial_t + v \partial_x) - \partial_x^2 +
 \frac{1}{\Lambda^2}\partial^4_x \right]\phi = 0.
\label{twaveeq}
\ea
When the flow is stationary, 
one can look for solutions with a fixed 
frequency $\la = i \partial_t$. 
Inserting $\phi = e^{-i\la t}\, \phi_{\la}(x)$ in \eqr{twaveeq} yields
\ba
\left[(-i \la + \partial_x v)( -i \la + v \partial_x) - \partial_x^2 +
 \frac{1}{\Lambda^2}\partial^4_x \right]\phi_\la = 0.
\label{waveeq}
\ea
Because of the quartic dispersion, the number of linearly independent solutions
is four. It would have been
$n$ if the dispersion relation would have been $\Om^2 = p^2 + p^n/\Lambda^{n-2}$ rather
than \eqr{disprela}. 
However, when imposing that the modes $\phi_{\la}$
be asymptotically bound for $x\to \pm \infty$, the dimensionality is reduced to $2$ or $1$
depending on whether $\la$ is real or complex, but irrespective of the value of the power $n$. 
(To avoid confusion about the real/complex character of $\la$
we shall write it as $\la= \om + i \Gamma$,  with $\om$ and $\Gamma$ both real and positive.
The other cases 
can be reached by complex conjugation and by multiplication by $-1$.)

%
The necessity of considering only 
asymptotically bound modes (ABM) comes from the 
requirement 
that the observables, such as the energy of \eqr{Hlindec}, be well defined. Returning to 
\eqr{action}, the conjugate momentum is \be
\pi = \partial_t \phi + v\partial_x \phi,
\label{pi}
\ee
the 
scalar product 
is 
\be
(\phi_1\vert\phi_2) = i\int_{-\infty}^{\infty}dx \left[\phi_1^*
\pi_2 
 - 
\phi_2 \pi_1^* 
\right] , 
\label{KGnorm}
\ee
and the
Hamiltonian is given by 
\ba
H = \frac{1}{2}\int dx \left[(\partial_t \phi)^2 + (1 - v^2) (\partial_x \phi )^2 + 
\frac{1}{\Lambda^2}(\partial_x^2 \phi)^2 \right] .
\label{Hphi}
\ea
In the subspace of ABM, the Hamiltonian is hermitian, i.e. $(\phi_1\vert H \phi_2 ) = (H \phi_1  \vert \phi_2)$.

We conclude 
 with some remarks.
First, when written in the form \eqr{KGnorm}, the scalar product is 
conserved in virtue of Hamilton's equations, and the hermiticity of $H$. 
Second, from \eqr{Hphi}, one sees that the Hamiltonian density is negative 
where the flow is supersonic: $v^2 > 1$. 
We shall later see that
the supersonic region should be `deep' enough so that it can sustain at least a bound mode,
thereby engendering a laser effect.
Third, when considering ABM, 
eigenmodes characterized by different frequencies are orthogonal in
virtue of the identity~\cite{Fulling:1989nb,Kang:1997uw}
\be
(\la' - \la^*)\, (\phi_\la \vert\phi_{\la'}) = 0,
\label{ortho}
\ee
which follows from 
the hermiticity of $H$. 
Finally, we notice that complex frequency ABM can exist  
in the present case because 
neither the scalar product \eqr{KGnorm}, nor the Hamiltonian \eqr{Hphi} is positive definite, 
see \cite{Fulling:1989nb} p. 228.
On the contrary, since fermionic fields 
are endowed with a positive scalar product~\cite{Itzykson:1980rh}, 
no complex frequency ABM could possibly be found in their spectrum~\cite{XXX}.

\section{The set of asymptotically bound modes}
\label{setofmodes}

\subsection{Main results}

In the BH-WH 
flows of 
\eqr{vdparam}, 
the set of ABM 
 contains a continuous spectrum of dimensionality $2$ labeled by
a {positive} real frequency $\om$, 
and a discrete spectrum of $N < \infty$ pairs of complex frequencies 
eigenmodes. 
We shall 
suppose 
that this set is complete.\footnote{We are currently trying 
to demonstrate this assumption.} 
That is,
any 
solution of \eqr{twaveeq} with $(\phi \vert\phi)  < \infty$ can be decomposed as
\ba
\phi 
&=& 
\int_0^\infty  d\om \left( e^{- i \om t}\, [ a_{\om, \,u}\, \phi_{\om}^u(x)  + a_{\om, \,v}\, \phi_{\om}^v(x)]     
+ h.c.\right)
\nonumber \\
&& + \Sigma_{a=1,N} \left( e^{- i \la_a t} \,  b_a \,\varphi_a(x) +  e^{- i \la^*_a t}\, c_a \psi_a(x) +  h.c.\right).
\label{lindec}
\ea
For flows
that are 
asymptotically constant 
on both sides of the BH-WH pair, 
we shall show that the real frequency modes can be normalized according to
\be
(\phi^i_{\om'}\vert\phi^j_\om) = \delta^{ij} \, \delta(\om - \om') ,\quad (\phi^{i\, *}_{\om'}\vert\phi^j_\om) = 0,
\label{delta_om}
\ee
where the discrete index $i$ takes two values $u,v$, and where $\om , \om' >0$.
The 
index $u,v$ 
characterizes modes which are asymptotically 
left ($v$) or right moving modes ($u$)
with respect to 
stationary frame.


When $\la$ is complex, the situation is unusual. Yet it closely corresponds to that 
described in 
the Appendix of \cite{Fulling:1989nb}. In fact whenever a hermitian Hamiltonian
possesses 
complex frequency ABM, 
one 
obtains 
a discrete set of two-modes $(\varphi_a, \psi_a)$ of complex conjugated frequency $\la_a, \la_a^*$.
Their `normalization' can be chosen to be
\be 
(\varphi_{a'} \vert \varphi_a) =0 ,\quad (\psi_{a'} \vert \varphi_a) = i\,  \delta_{a, a'}, 
\label{psivar}
\ee
with all the other (independent) products vanishing in virtue of \eqr{ortho}. 
Since the overlap
between modes belonging to the continuous and discrete sectors, such as $(\phi_{\om}\vert\varphi_a)$, also vanish,
eigenmodes of 
different frequency never mix. Moreover, 
since the positive norm modes $\phi^i_\om$ all have $\om > 0$, 
one cannot obtain
Bogoliubov transformations as those characterizing the 
Hawking radiation associated with 
a single BH (or WH). 
This implies that the (late time) radiation emitted by BH-WH pairs 
{\it entirely} comes from 
the discrete set of modes.

Using the above equations, the energy carried by $\phi$ of \eqr{lindec}
is given by 
\be
E = (\phi\vert H \phi)= 
 \int_0^\infty  d\om\,  \om \left( \vert a_{\om, \,u}\vert^2 + 
\vert a_{\om, \,u}\vert^2 \right)
 + \Sigma_{a=1,N} \left( -i \la_a \,  b_a c_a^* +  h.c.\right) .
\label{Hlindec}
\ee
Because of the complex frequency modes, it is unbound from below. 
Notice also that the absence of terms such as $\vert b_a\vert^2$
is necessary to have at the same time complex frequency eigenmodes
and real energies. 

\subsection{Asymptotic behavior and roots $k_\la$}

The material presented below closely follows that of \cite{Macher:2009tw}. 
In fact the lengthy presentation of that work was written having in mind its applicability
to the present case.   
The novelties are 
related to the fact that, for the metrics of \eqr{vdparam},
the supersonic region is bound (from $-L$ to $L$), and the
velocity is subsonic for $x\to \pm \infty$.

Since the velocity $v$ is asymptotically constant for $|x/L| \gg 1$, 
in both asymptotic regions,
the solutions of Eq.~(\ref{waveeq}) are superpositions of four exponentials 
$e^{i k_\la x}$ weighted by constant amplitudes. 
To characterize a solution, one thus needs to know (on one side, say on the left) 
the four amplitudes $A_k$ 
associated to the corresponding
asymptotic wave vectors $k(\la)$. 
These are the 
roots of 
\be
(\la - v({x}) k )^2  = k^2 +  \frac{k^{4}}{ \Lambda^{2 }} = \Omega^2(k),
\label{reldisp}
\ee
evaluated for $v(x \to - \infty)= v_-$. 
%
We shall not assume {\it a priori} that $\la$ is real. 
Rather we shall look for all ABM. 
Notice that when considering complex frequencies $\la= \om + i \Gamma$, 
the roots of \eqr{reldisp} are 
continuous functions of 
$\Gamma$. 
In addition, when the scales are well separated, i.e. 
when $\kappa/\Lambda \ll 1$, the relevant values of $\Gamma$ will obey $\Gamma/\Lambda\ll 1$.
It is therefore appropriate to 
start the analysis with $\la= \om$ real, and then to study how the roots migrate when 
$\Gamma$ increases.

When $\om > 0$,  since the flow is subsonic for $x \to \pm\infty$, 
there exist two real asymptotic roots:  
$k^u_\om > 0$ and $k^v_\om <  0$ which correspond to a right and 
a left mover respectively. 
There also exist a pair of complex conjugated roots, since \eqr{reldisp} 
is real.
Thus, on each side of the BH-WH pair there is a growing and a decaying mode. As in \cite{Macher:2009tw} we define 
them 
according to 
the 
behavior of the mode when moving away from the BH-WH horizons.

\begin{figure} 
\includegraphics[scale=0.3]{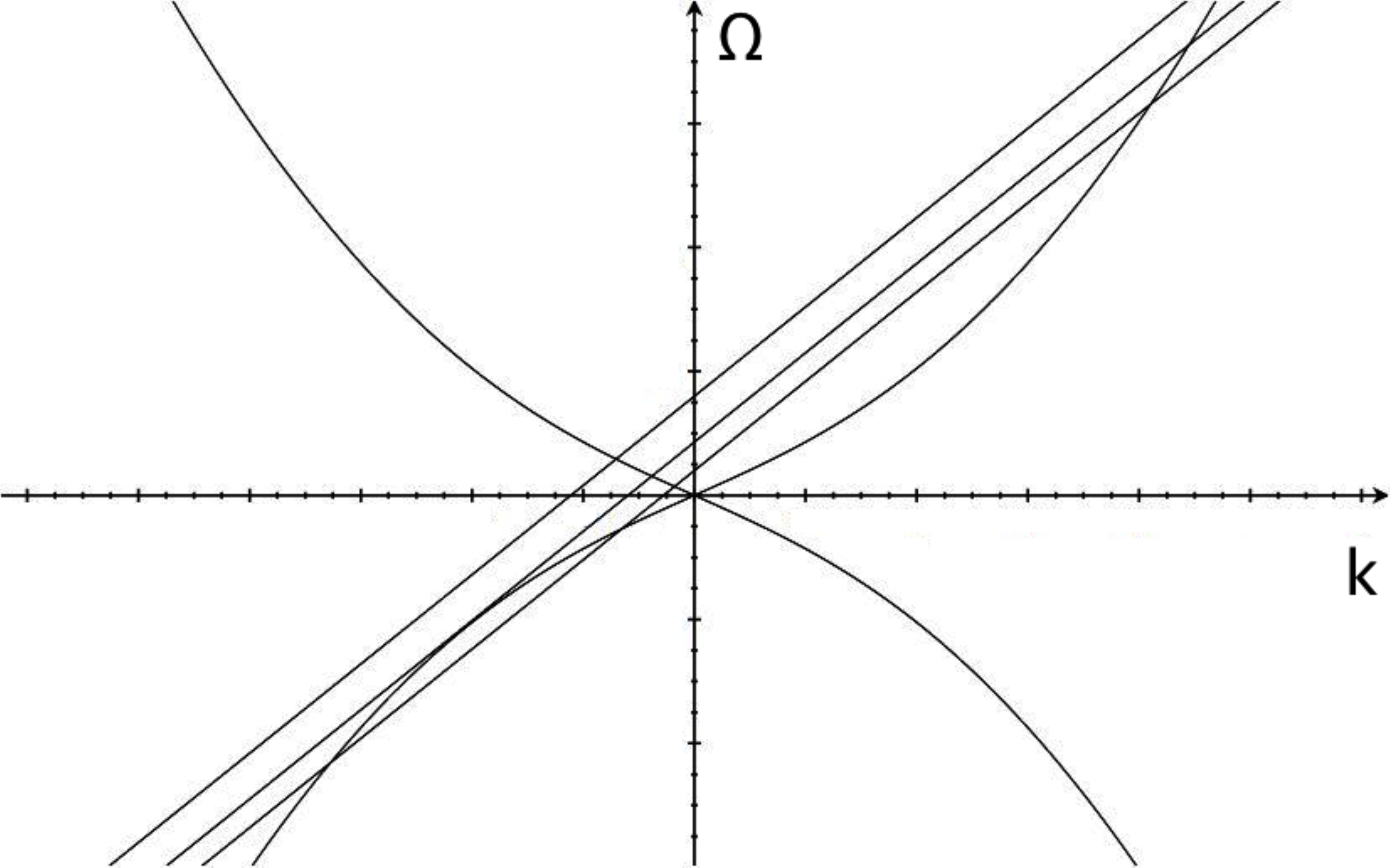}
\includegraphics[scale=0.3]{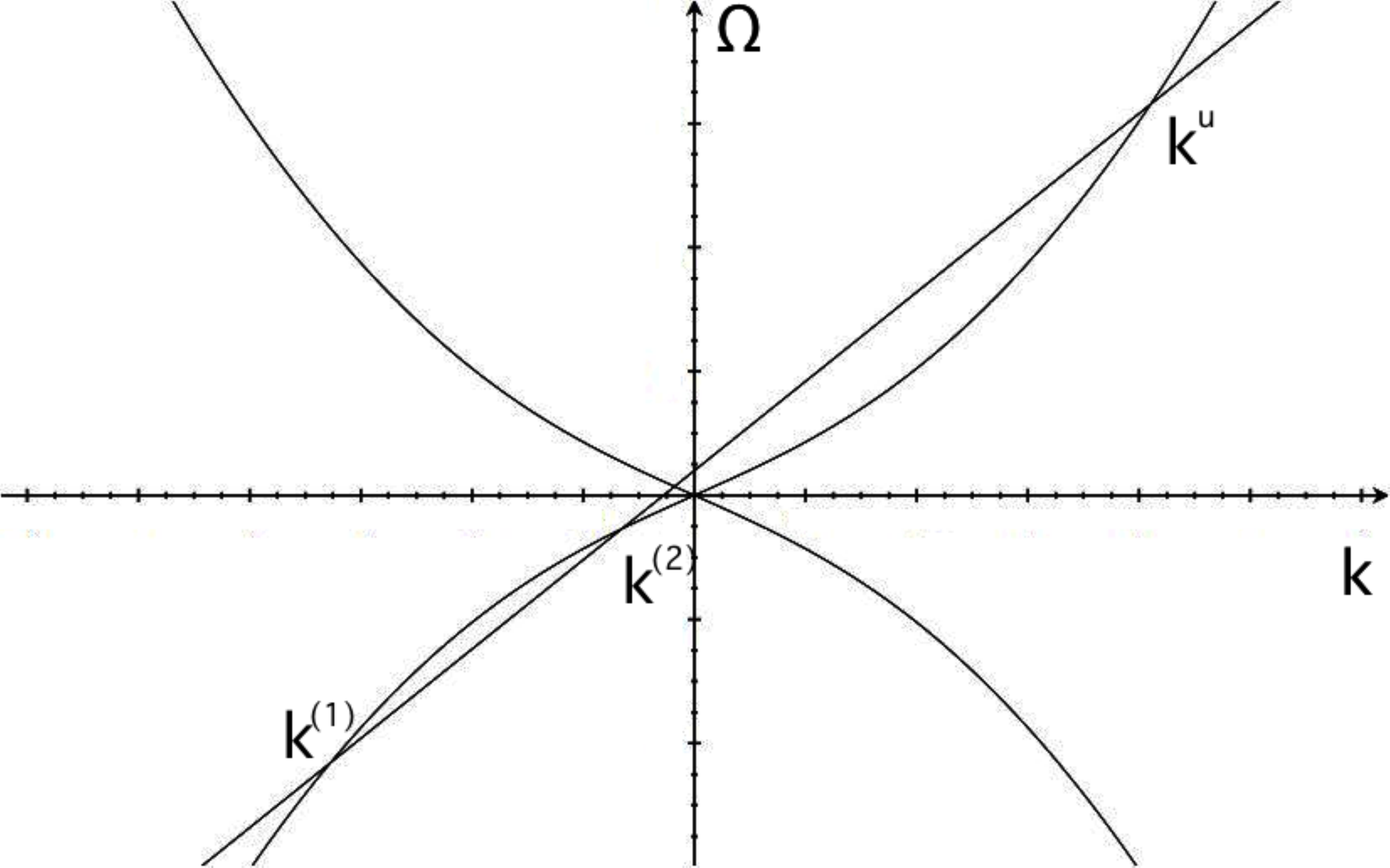}
\caption{Graphical resolution of \eqr{reldisp}.
On the left, the straight lines represent $\om - k v_+$ in the supersonic regime $v_+ < -1$
for three different values of $\om$. The middle one, which is tangent to $\Om(k)$,
determines the critical frequency $\om_{\rm max}$. For $\om < \om_{\rm max}$, the two extra
negative roots correspond to right moving modes with respect to the fluid since $\partial_k \Om > 0$. On the right,
 $k^u$ is the positive root describing the right moving mode, see \eqr{WKBmodes},
$k^{(1)}$ the most negative root, $k^{(2)}$ that with the smallest norm, see \eqr{phiuin}. 
}
\label{figdisp} 
\end{figure}

In preparation for what follows, 
we study
the roots in the supersonic region between the horizons.
For $\om$ 
smaller than a critical frequency $\om_{\rm max}$, whose value is discussed below,
the four  roots are real. 
For flows to the left, $v < 0$, 
the two new real roots correspond to two right movers (with respect to the fluid).
Indeed, they live on the $u$ branch of the dispersion relation \eqr{reldisp}, 
that with $\partial_k \Omega > 0$, see Fig.\ref{figdisp}.
When $\om$ increases at fixed $\vert v \vert> 1$, these roots approach to each other.
Thus for flows characterized by a maximal velocity $v_+$, 
there is a frequency $\om_{\rm max}$ above which they 
no longer exist as real roots. It is given
by the value of $\om$ where they 
merge 
for $v = v_+$. 
When the two horizons are well separated, $v_+$ is given in \eqr{v+},
up to exponentially small terms.
As shown in \cite{Macher:2009tw},  
$\om_{\rm max}$ 
is of the form $
\om_{\rm max} = \Lambda f(D)$, where $D$ is defined in \eqr{vdparam}.
For $D\ll 1$, one finds $f(D)\propto D^{3/2}$. 
Thus, for a given 
dispersion scale $\Lambda$, 
$\om_{\rm max}$ can be arbitrarily small. 
This will be important when considering the appearance
of the laser effect in parameter space. 

\subsection{The continuous spectrum}

We now have all the elements to show that the continuous part 
of the spectrum of ABM is labeled by positive real frequencies $\om$, 
and that, for a fixed $\om$, its 
dimensionality 
is $2$. 
The general solution of \eqr{waveeq} with $\om$ real 
can be characterized by the $4$ 
amplitudes, which multiply
the four exponentials 
evaluated in the asymptotic left region.
When imposing that the growing mode is absent on that side,
only three independent solutions remain. However, when propagating 
these solutions in the asymptotic
region on the right side of the BH-WH pair, 
the growing mode will be generally present on that side. 
Thus when requiring that it be also absent imposes to take particular combinations, 
and this reduces the dimensionality of ABM 
to two. 

One can then take 
appropriate linear combinations 
to construct the $in$ ($out$) modes describing the left and right movers
propagating toward (escaping from) the BH-WH pair. 
The $in$ right moving solution $\phi_\om^{u, \, in}$ is the combination which on the left 
is asymptotically proportional to $e^{i k^u_\om x}$ where $k^u_\om$ is the asymptotic
real positive root of \eqr{reldisp}. Similarly one can identify the $in$ left moving mode $\phi_\om^{v\, in}$,
and the two $out$ modes $\phi_\om^{u\, out}$ and $\phi_\om^{v\, out}$. 

More precisely, because of the infinite and flat character of the space on either side of the BH-WH pair, the
two $in$ and the two $out$ modes can be normalized as in a constant velocity flow 
(by considering a series of broad wave packets localized in one asymptotic region 
and whose spread in frequency progressively vanishes).
Considering $\phi_\om^{u\, in}$ for ${x \to -\infty}$, \eqr{delta_om} and \eqr{KGnorm} 
imply that it asymptotes to 
\be
\phi^{u\, in}_{\om}(x) \to \sqrt{\frac{dk_\om^u}{d \om}}\,  
\frac{\exp{i k^u_\om x}}{\sqrt{4 \pi  \Omega(k^u_\om)}}.\label{normmodes}
\ee
A similar expression valid for $x\to \infty$ gives $\phi^{v\, in}_{\om}$.
These two $in$ modes 
are 
orthogonal to each other, establishing the Kronecker $\delta^{ij}$ in \eqr{delta_om}. 
For $\om > 0$, these modes have a positive KG norm.
For $\om < 0$, they  have a negative norm. Therefore 
negative norm modes can all be described 
as superpositions of complex conjugated positive norm modes. 

When propagated 
across the BH-WH geometry, the $in$ modes are scattered by the
gradients of $v(x)$. 
When the variation of $v$ is slow, i.e. $\partial_x \ln v \ll \partial_x \ln k_\om$,
the exact solutions are globally well approximated by
the WKB solutions of \eqr{waveeq}. For the $u$-mode, one finds, see \eqr{normmodes},
\be
\varphi^{u}_{\om}(x) = \sqrt{\frac{\partial k_\om^u(x)}{\partial \om}}\,  
\frac{\exp \left(i \int^x_{-L} dx' k^u_\om(x')\right)}{\sqrt{4 \pi  \Omega(k^u_\om(x))}}.\label{WKBmodes}
\ee
We use the symbol $\varphi^u_\om$ ($\varphi^v_\om$)
to differentiate the 
WKB solution from the exact one $\phi^u_\om$ ($\phi^v_\om$). 
At high frequency, $\om/\kappa \gg 1$, the inequality $\partial_x \ln v \ll \partial_x \ln k_\om$
is satisfied, and 
 $u$ and $v$ modes do not mix. 
At lower frequency, they do. Nevertheless, far away from the BH-WH pair, since $v(x)$ 
 is asymptotically constant, exact solutions 
decompose into superpositions of $\varphi^u_\om$ and $\varphi^v_\om$ with constant amplitudes.
This applies for both the real and the complex frequency modes of \eqr{lindec}.
We shall use this fact several times to characterize the properties of the exact solutions.

Introducing the 
$out$ modes as in \eqr{normmodes}, 
this scattering 
is 
described by 
\be
\begin{split}
\phi^{u, in }_{\om} &= T_{\om} \, \phi^{u, out}_{\om} 
+ R_{\om} \,  \phi^{v, out}_{\om},\\
\phi^{v, in }_{\om} &= \tT_{\om}\,  \phi^{v, out}_{\om} + \tR_{\om}\,  \phi^{u, out}_{\om} .  
\end{split}
\label{RT}
\ee
Unitarity imposes $| T_{\om}|^2 +|R_{\om}|^2 = 1 = | \tT_{\om}|^2 +|\tR_{\om}|^2 $, 
and $R_\om \tT_\om^* + T_\om \tR_\om^* = 0$. 
For all values of $\om$, one thus has an elastic scattering, 
without spontaneous pair creation. For frequencies $\om > \om_{\rm max}$,
 \eqr{RT} coincides with what is found in single horizon scatterings~\cite{Macher:2009tw}.
Instead, for frequencies $0 < \om < \om_{\rm max}$,
this radically 
differs from the scattering on a single 
because, in that case, 
the 
matrix was $3 \times 3$
and mixed $\phi^u_\om$, $\phi^v_\om$
with the negative frequency 
$u$-mode $(\phi_{-\om}^{u})^*$. 
The presence of the second horizon therefore `removes' 
these 
modes. 
As we shall later see, they shall be `replaced' by a finite and discrete set of complex frequency modes.
This is not so surprising since the classical trajectories associated with the 
negative frequency modes are closed (hence the discretization), and since these trapped
modes mix with the continuous spectrum through each horizon (hence the imaginary part of the frequency).
In fact this is reminiscent, but not identical, to 
quasinormal modes~\cite{Frolov:1998wf} or resonances. 
The main difference is that quasinormal modes are not asymptotically bound.
Thus, they should not be used in the mode expansion of \eqr{lindec}.

It should be also mentioned that both 
$in$ and $out$ modes contain a trapped component 
in the supersonic region\footnote{Because of this trapped wave, 
the $in$ (and $out$) modes are {\it not} asymptotic
modes in a strong sense since a wave packet made with $\phi_\om^{u,\,  in}$
will have a double spatial support for $t\to -\infty$:
the standard incoming packet coming from $x= -\infty$, and the unusual trapped piece. 
This additional component, see \eqr{phiuin}, ensures 
that $\phi_\om^{u,\,  in}$ 
is orthogonal to the complex frequency modes of \eqr{lindec}.}
which plays no role as far as their normalization 
is concerned
since the modes are everywhere regular
and 
the supersonic domain 
is finite.
The case where the subsonic domain is also finite should be analyzed
separately. 
If periodic conditions are imposed at the edges of the condensate, 
one obtains discrete frequencies and resonances 
effects~\cite{Garay:2000jj,Gardiner}.
If instead absorptive conditions are used, the frequencies are continuous
and the situation is closer to the case we are studying.

\subsection{The discrete spectrum}
\label{discrete}

On general grounds we explain why, when considering \eqr{waveeq} in BH-WH flows of 
\eqr{vdparam}, there exists
a discrete and finite set of complex frequency modes. 
To this end, we first show that for a generic complex frequency $\la = \om + i \Gamma$,
there exists no ABM.
%
When 
$\Gamma\ll \Lambda$, the four 
roots have not crossed each other with respect to the case where $\Gamma = 0$ 
since the imaginary part of the two complex roots $k^\pm_\om$ 
with $\om$ real is proportional to $\Lambda$. 
Thus, in this regime, 
one can still meaningfully
talk about the two 'propagating' roots $k^u_\la, k^v_\la$, and the growing and decaying roots $k^\pm_\la$.  
Then, as in the former subsection, 
when imposing that the growing mode be asymptotically absent on both sides of the BH-WH pair,
the space of 
solutions is still two. 
However when $\Gamma > 0$, the $u$-root $k^u$ acquires a positive imaginary contribution,
which means that the $u$-mode of \eqr{normmodes} diverges for $x \to -\infty$.
To get a bound mode, its amplitude should be set to $0$.
For similar reasons,
on the right, the incoming $v$ mode diverges for $x \to \infty$. Hence, for a general value of $\la$,
the set of ABM 
is empty.
The  
above reasoning 
applies to flat backgrounds with $v$ constant, and 
 establishes that in that case, complex frequency modes should not be considered in 
\eqr{lindec}. 

We should now explain why when $v$ is supersonic in a finite region, 
some complex frequency ABM exist. 
The basic reason is
the same as that which gives rise to a discrete set of bound modes 
when considering the Schroedinger equation in a potential well (or the propagation of light through a cavity).
In the supersonic region, (in the well), there exist two additional real roots $k_\om$ when $\la = \om$ real.
The classical trajectories associated with them 
are closed, and, 
as in a Bohr-Sommerfeld treatment, the discrete set of 
modes is 
related to the requirement that the mode be single valued and bound.
%
The complex character of the frequency $\la$ is due to 
the finite tunneling amplitude 
across the horizons. Indeed, as we shall later see, 
the imaginary part of $\la_a$ is proportional to some quadratic expression in
the $\beta$ Bogoliubov coefficients characterizing the scattering through the horizons. 
Were these 
coefficients equal to zero,  $\la_a$ would have been real.
When `turning on' these coefficients, 
the frequency $\la_a$ migrates in the complex plane, 
and the bound modes are continuously deformed.  

These ABM 
appear in pairs with complex conjugated frequencies. 
This stems from the hermiticity of $H$ which guarantees that there exists an ABM 
of frequency $\la_a^*$ whenever there is one of frequency $\la_a$.
At this point it should be re-emphasized that the existence of these complex frequency ABM
is due to the fact that the scalar product is not definite positive.
Indeed, these modes all have a vanishing norm
in virtue of \eqr{ortho}. 
We can also conclude that the discrete set of complex frequencies ABM is 
finite. In fact there are no closed orbits 
for 
$\om < -\om_{\rm max}$, since the extra real roots $k_\om$
no longer exist 
 and since there is a gap between the  eigenfrequencies.

\subsection{The quantization}

The canonical quantization of the field $\phi$ is straightforward since each eigenfrequency sector evolves 
independently from the others. 
Indeed when decomposing the field as in \eqr{lindec}, with the 
coefficients $a, b, c$ promoted operators,
the equal time commutation $[\phi(t,x),\pi(t,x')] = i \delta(x-x')$ and the orthonormality conditions
\eqr{delta_om} and \eqr{psivar} entirely fix their commutation relations. 
For real frequency modes, the operators $a_{\om,\, i}= (\phi_\om^i\vert \phi)$ 
obey the standard commutation relations
\be
[a_{\om,\, i},\, a_{\om,\, j}^\dagger ]= \delta_{ij} \, \delta(\om - \om').
\ee
Instead, for complex frequency modes, one gets 
\be
[b_{a},\, c_{a'}^\dagger ]= i \delta_{a\, a'}.
\label{combc}
\ee
All the other commutators vanish.

Because of this disconnection, the ground state of the real frequency modes is
stable and in fact subject to no evolution. Hence the number of quanta of these modes
is constant. 
The evolution of the states associated with the complex frequency modes is also rather simple
and described in the Appendix.
What remains to be done is 
to determine the properties of the asymptotic fluxes. To this end we need 
a better understanding of the 
modes. 

\section{The properties of the modes}
\label{modeprop}

From 
\eqr{waveeq}, it is not easy to determine the 
complex frequencies $\la_a$ and the properties of the modes.
Several routes can be envisaged. One can adopt numerical techniques. 
We are presently modifying~\cite{Finazzi} the code used in \cite{Macher:2009tw,Macher:2009nz} 
to address this 
problem. 
One can also bypass the calculation of the eigenmodes and directly compute
the propagation of coherent states, or the density-density correlation function~\cite{Balbinot:2007de},
using the techniques of~\cite{Carusotto:2008ep}. This is currently under study~\cite{Iacopo}.
One can also envisage to use analytical methods by
choosing the 
flow $v(x)$ as in  
\cite{Barcelo:2006yi}. 
This method is also currently under study~\cite{Carlos}.

In what follows, 
we use an approximative treatment which is valid 
when the two horizons are well separated. 
Doing so, we make contact with the original treatment~\cite{Corley:1998rk,Leonhardt:2008js} based on wave packets.
More importantly, we determine algebraic relations 
which do 
not rely on the validity of our approximations. 
In particular,
 we establish that 
the real frequency modes $\phi_\om^u$ are intimately related to 
the 
complex frequency modes even though their overlap vanish.

\subsection{The limit of thin near horizon regions}

To simplify the mode propagation, 
we assume that the near horizon regions are thin and well separated, i.e. $L \kappa \gg D$ in \eqr{vdparam}. 
In this case, the propagation through the BH-WH geometry resembles very much to that through a cavity.
Indeed, the following 
apply. 
First, the nontrivial propagation across the two thin horizon regions
can be described by 
matrices that connect a solution evaluated on one side to that on the other 
side.
Second,
the 
modes
can be analyzed separately in three regions: in $L$, the external left region, for $-(x + L) \gg  D/\kappa_W$;
 in $R$, the  external right region, for $(x - L) \gg  D/\kappa_B$; 
and in the inside region $I$, 
for $L - \vert x \vert \gg D/\kappa$. 
Within each region, the gradient of $v$ is small. Hence, any solution is well approximated 
by a superposition of WKB waves \eqr{WKBmodes} with constant amplitudes.

To further simplify the analysis, we use the fact that 
the $u$-$v$ mode mixing coefficients are generally much smaller than those mixing the negative frequency
modes to the positive $u$ ones~\cite{Macher:2009tw,Macher:2009nz}. Hence, it is a reliable (and consistent) 
approximation to assume that the $v$ modes completely decouple. After having analyzed this case,
we shall briefly present the modifications introduced by relaxing this hypothesis.
Adopting the hypothesis that the $u-v$ mixing can be neglected, 
for each $\om$ real, one has the following situation. In the left region $L$, 
one only has the WKB mode $\varphi^u_\om$ of \eqr{WKBmodes}. Thus, the only solution is $\phi^{u,\, in}_\om$ (up to an overall irrelevant phase we take to be $1$).
In the inside region $I$,
one has three modes: 
\be
\phi_\om^{u, \, in} = {\cal A_\om} \, \varphi^u_\om + 
{\cal B}^{(1)}_\om (\varphi^{(1)}_{-\om})^* + 
{\cal B}^{(2)}_\om (\varphi^{(2)}_{-\om})^* ,
\label{phiuin}
\ee
since in supersonic flows, 
there exist two extra real 
roots in \eqr{reldisp}. 
The superscripts $u, (1), (2)$ characterize the coefficients and the WKB  modes associated with 
the three roots shown in Fig. \ref{figdisp}. 
Since we are considering a 
solution with $i \partial_t = \om > 0$, 
the (positive norm) negative frequency modes $\varphi_{-\om}^{(i)}$
appear complex conjugated in \eqr{phiuin}.

In the external $R$ region the solution must be again proportional to 
the WKB mode $\varphi^u_\om$. 
By unitarity, 
the solution must be of the form $\phi_\om^{u, \, in} = e^{i \theta_\om}\varphi_\om^{u} $.
Thus, a full characterization of 
$\phi^{u,\, in}_\om$
requires to compute the phase $\theta_\om$ and the above three coefficients. 
At this point, it should be noticed that 
$\la = \om$ is {\it a priori} real. 
However 
the $S$ matrix, and therefore the three coefficients, are holomorphic
functions in $\la$. Hence nothing prevents to leave the real axis. In fact we shall show
that the complex frequencies 
correspond to poles associated with a coefficient of the $S$ matrix. 

\subsection{An $S$ matrix approach}
\label{Sma}

\subsubsection{The $S$ matrix}

The simplest 
way we found to compute the above coefficients is
to 
follow the 
 approach  of \cite{Leonhardt:2008js}, up to a certain point.
In this treatment, a solution of frequency $\om > 0$ 
is represented by a two component vector $(\varphi^u_\om, \varphi^*_{-\om})$.
The time evolution of a wave packet of such solutions is then considered 
in the thin horizon limit.
Since the frequency content of the wave packet plays no role, 
we do not need to introduce a new notation to differentiate it from an eigenmode. 
In this language, the $S$-matrix characterizing a bounce of the trapped mode  $\varphi^*_{-\om}$
can be decomposed as 
\be
S = U_4 \,U_3 \,U_2 \,U_1 .
\label{Utot}
\ee
The first matrix describes the scattering across the WH horizon. In full generality 
we parameterize it by
\ba
U_1 =  \left(\begin{array}{cc}\alpha_\om &\alpha_\om\, z_\om 
\\ \tal_\om \,  z_\om^*  & \tal_\om \end{array}\right)
\ea
Unitarity imposes that $\vert \alpha_\om\vert^2 = \vert \tal_\om\vert^2$ and
$\vert \alpha_\om\vert^2\, (1 - \vert z_\om\vert^2 ) = 1$.
The second matrix describes the free propagation (i.e. without backscattering) 
from the WH to the BH horizon 
\ba
U_2 = \left(\begin{array}{cc} e^{iS^u_\om} & 0  \\
     0 & e^{-iS^{(1)}_{-\om}}
\end{array} \right) .
\label{U2}\ea
In a WKB approximation, the two phases are respectively given by the actions
\ba
S^u_\om &=& \int_{-L}^L dx \, k^u_\om(x), \nonumber\\
S^{(1)}_{-\om} &=& 
 \int_{L_\om}^{R_\om} \! dx \left[ - k^{(1)}_{\om}(x)\right],
\ea
In $U_2$,  
$S^{(1)}_{-\om}$ is multiplied by $-i$ since it governs the evolution of $\varphi_{-\om}^*$.
Its momentum is $k^u_{-\om} = - k^{(1)}_{\om} > 0$, where $k^{(1)}_{\om}$
is the most negative root of \eqr{reldisp} found in supersonic flows. The ends of integration 
$L_\om$, $R_\om$ are, respectively, the left and right locations of the turning
points of the trajectories with 
 $-\om < 0$. These 
obey 
Hamilton's equations~\cite{Brout:1995wp,Balbinot:2006ua}
\be
dx/dt = (\partial_\om k^u_\om)^{-1}, \quad dk^u/dt = - \partial_x \om = - k^u \partial_x v ,
\ee
for negative frequency.
In the thin horizon approximation,
the turning points hardly differ from $-L$ and $L$. 
Unitarity brings no conditions on these phases.

The third matrix describes the scattering across the WH horizon and we write it as 
\ba
U_3 = \left(\begin{array}{cc}\gamma_\om &\gamma_\om \, w_\om 
\\ \tga_\om \, w_\om^*  & \tga_\om \end{array}\right).
\label{U3}
\ea
Unitarity imposes that $\vert \gamma_\om\vert^2 = \vert \tga_\om\vert^2$
and $\vert \gamma_\om\vert^2 \, (1 - \vert w_\om\vert^2 ) = 1$.
The fourth matrix describes the return of
the negative frequency partner towards the WH horizon, 
whereas the positive frequency mode propagates away in the $R$ region.
 This is described by 
\ba
U_4 = \left(\begin{array}{cc} 1 & 0  \\
    0 & e^{iS^{(2)}_{-\om}} \end{array} \right) .
\label{U4}\ea
In the WKB approximation, 
 this backwards movement (hence the $+i$ in the front of $S^{(2)}$) 
is governed by 
\ba
S^{(2)}_{-\om} =
\int^{R_\om}_{L_\om} dx [-k^{(2)}_{\om}(x)],
\ea
where the momentum $k^{(2)}_{\om}$ is the least negative 
$u$-root of \eqr{reldisp}. 
Since the positive frequency mode further propagates to the right, 
there is no meaning to attribute it a phase in $U_4$. 
In any case this phase would drop from all physical quantities.

The 
matrix $S$ of \eqr{Utot} is 
unitary since its 4 constituents are.
Hence $\vert S_{22} \vert^2 = \vert S_{11} \vert^2 = 1+  \vert S_{12} \vert^2 $. 
The components $S_{22}$ and $S_{21}$ we shall later use are given by
\ba
S_{22} &=&  \tga_\om  \tal_\om \, e^{-i(S^{(1)}_{-\om} -S^{(2)}_{-\om})}
 \left(1 +  z_\om w^*_\om  \frac{\alpha_\om}{\tal_\om }\, e^{i(S^{u}_{\om}+S^{(1)}_{-\om})} \right), 
\nonumber \\ 
S_{21} &=& \tga_\om  \tal_\om \, e^{-i(S^{(1)}_{-\om} -S^{(2)}_{-\om})}
\left( z_\om^* + w^*_\om  \frac{\alpha_\om}{\tal_\om }\, e^{i(S^{u}_{\om}+S^{(1)}_{-\om})} \right) .
\label{S22}
\ea
Hitherto we 
followed the method of \cite{Leonhardt:2008js}.
Henceforth, we proceed differently by adding a key element. 
We require that the mode propagated by $S$ be single valued. 
For real $\om$, this unequivocally defines 
$\phi^u_\om$ of \eqr{phiuin}. 
Moreover, when looking for complex frequency 
bound modes, this will give us the modes $\varphi_a,\psi_a$ we are seeking. 

\subsubsection{The real frequency modes} 

Imposing that the trapped mode of negative frequency 
is single valued 
translates into 
\ba
\left(\begin{array}{c} e^{i \theta_\om}\\ b_\om \end{array}\right)
= 
\begin{array}{cc} S \end{array}
\left(\begin{array}{c} 1 \\ b_\om \end{array}\right).
\ea
The phase $\theta_\om$ is that mentioned after \eqr{phiuin}. It
should not be constrained since the 
positive frequency component keeps
propagating to the right. 
The 
matricial equation gives
\ba
b_\om &=&{\frac{ S_{21} }{1-S_{22}}}  , \nonumber \\
e^{i \theta_\om} &=& - \frac{S_{11}}{S_{22}^* } \frac{ 1 - S_{22}^* }{1-S_{22}}.
\label{Unotyetpole}
\ea
These equations constitute the first important result of this section. 
They 
do not rest on the WKB approximation. 
Of course, this approximation can be used to estimate the 4 elements of $S$.
But once these are known, e.g. using a numerical treatment, these equations apply. 
What is needed to get these equations is the neglect of the $u-v$ mixing,
and no 
significant frequency mixing 
in the inside region in order to obtain well-defined 
amplitudes 
in \eqr{phiuin}.

Using $S^\dagger = S$ one verifies that the norm of the right hand side of 
the second equation is unity. 
This is as it is must be, since in the absence of $u-v$ mixing,
the $u$-component only acquires a phase, here measured with respect to 
the WKB wave. 
Adopting the convention that WKB modes $\varphi_\om$ have a vanishing phase at $x=-L$, 
the coefficients of \eqr{phiuin} are
\ba
{\cal A_\om} &=&  \alpha_\om (1 + z_\om b_\om) ,
\nonumber \\
{\cal B}^{(1)}_\om &=& \tal_\om (z^*_\om + b_\om), 
\quad \, 
{\cal B}^{(2)}_\om =  b_\om .
\label{bB}
\ea
These amplitudes are governed by $b_\om$, 
which 
can {\it a priori} 
be larger or smaller than unity.
In particular it diverges 
if $S_{22} \to 1$ for some 
$\om$,  
thereby approaching a resonance,
see Fig. 3.
We now show that the complex frequency modes 
correspond to these resonances. 

 \begin{figure}  
\centering
\includegraphics[height=70mm]{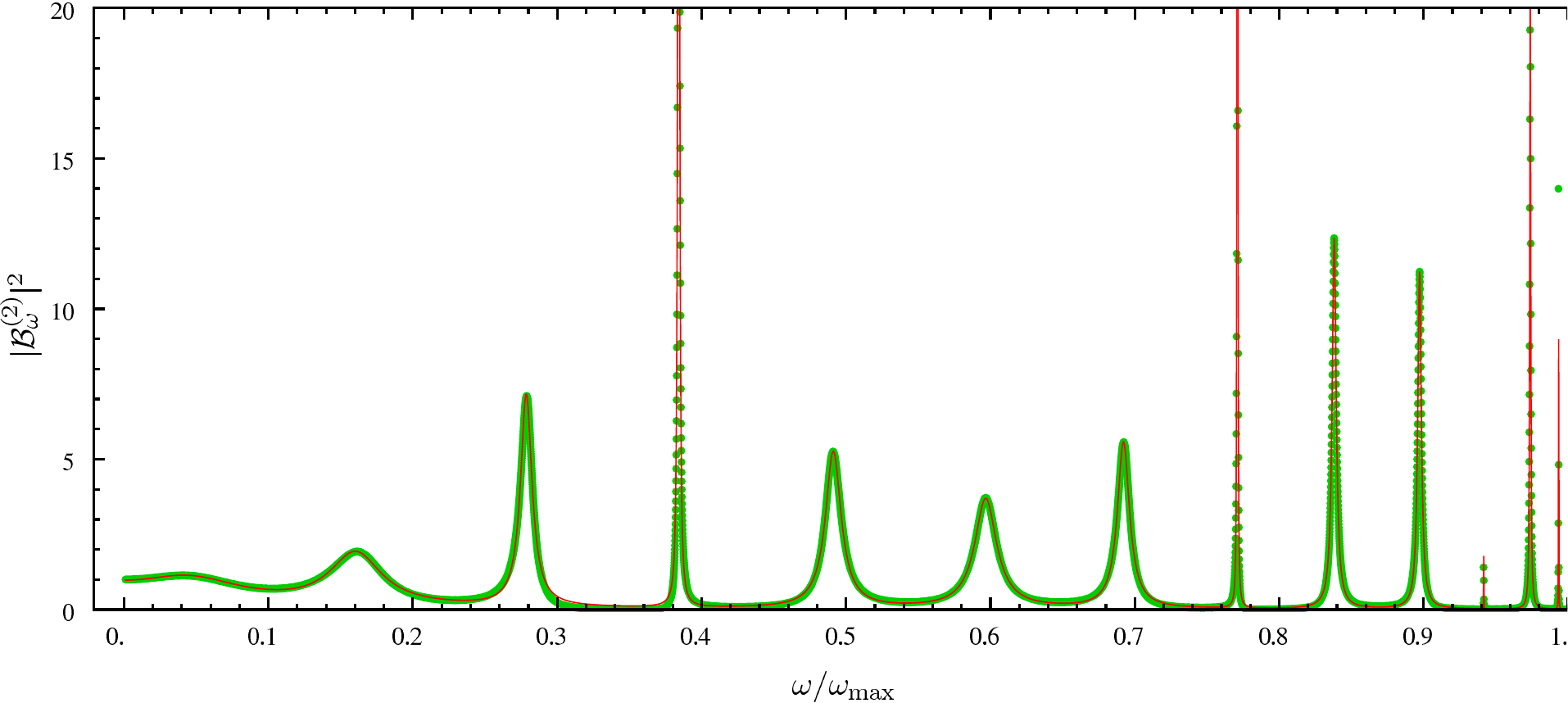}
\caption{
By making use of numerical results borrowed from~\cite{Finazzi}, 
we have represented the square norm of ${\cal B}^{(2)}_\om$, the 
amplitude of the  negative frequency trapped mode of \eqr{phiuin}, as a function of $\om$ real.
Near a complex frequency $\la_a=
\omega_a - i\Gamma_a$ which solves $S_{22}=1$, see \eqr{Upole}, 
$\vert {\cal B}^{(2)}_\om \vert^2 $ behaves as 
 a Lorentzian:
$ 
\sim \left| {\omega - \omega_a - i\Gamma_a} \right|^{-2} $, see Eqs. (\ref{Unotyetpole}, \ref{bB}). 
The dots are the numerical values, whereas the continuous line is a fitted sum of Lorentzians. The remarkable agreement establishes that 
the complex frequencies $\la_a$ 
can be deduced from the analysis of the 
real frequency modes. The frequency $\om$ has been expressed in the units of 
$\om_{\rm max}$, see Fig. 2, so that there is no resonance above $\om/\om_{\rm max} = 1$. In the present case
there are 13 resonances. The narrow peaks, $\Gamma_a \sim 0$, are due to the fact that 
the surface gravities are equal, $\kappa_B= \kappa_W$,  
which leads to $z_\om= w_\om$ in 
\eqr{Gddr}.} 
\end{figure}

\subsubsection{The pairs of complex frequency modes} 

Following the 
discussion of Sec.\ref{discrete}, 
we impose that the amplitude of the incoming $u$ branch be zero, 
and, as above, that the trapped mode of negative frequency is single valued.
This  gives
\ba
\left(\begin{array}{c} \beta_a \\ 1 \end{array}\right)
= 
\begin{array}{cc} S \end{array}
\left(\begin{array}{c} 0 \\ 1 \end{array}\right),
\label{Uc}
\ea
which implies 
\ba
\beta_a = S_{12}, \quad 
1 =  S_{22}.
\label{Upole}
\ea
Two important lessons are obtained.
First, to get the complex frequencies $\la_a$ with a positive imaginary part, 
it suffices to solve the roots of $S_{22}=1$. 
Second, as mentioned, these correspond to the poles 
characterizing the propagating modes, see \eqr{Unotyetpole}. 
%

Before computing these frequencies, 
we explain how to get 
the decaying 
 modes $\psi_a$, the `partners' of the $\varphi_a$ in \eqr{lindec}. 
Since these bound modes have a negative imaginary frequency, 
the amplitude of the escaping mode must be zero, i.e.
\ba
\left(\begin{array}{c} 0 \\ 1 \end{array}\right)
= 
\begin{array}{cc} S \end{array}
\left(\begin{array}{c} \tilde \beta_a \\ 1 \end{array}\right).
\ea
Using the hermitian conjugated $S^\dagger$, and the unitarity relation $S^\dagger S = 1$, this condition gives
\ba
\left(\begin{array}{c} \tilde \beta_a\\ 1 \end{array}\right)
= 
\begin{array}{cc} S^\dagger \end{array}
\left(\begin{array}{c} 0 \\ 1 \end{array}\right),
\label{eqfpsi}
\ea
from which we get $\tilde \beta_a = (S^\dagger)_{12}$ and $1 =  (S^\dagger)_{22}$. 
As explained in the Appendix, when expressed in terms of the elements of $S$, 
these equations give  
\ba
\tilde \beta_a = - [S_{21}(\lambda^*)]^* , \quad 
1 =  [S_{22}(\lambda^*)]^*.
\label{Ugpole}
\ea
Thus the 
solutions of $1 =  S_{22}^*$ 
are the complex conjugated of those that solve \eqr{Upole}, 
thereby establishing the partnership between $\varphi_a$ and $\psi_a$.
In addition, one has $\tilde \beta_a = -(\det S)^*\beta_a$.

\subsection{The set of complex frequencies $\la_a$}
\label{zsmall}

To 
compute the roots 
of \eqr{Upole} we need to know $S_{22}$ 
 as a function of $\la= \om + i \Gamma$. 
In what follows 
we shall relate them to the quantities which enter 
in  \eqr{S22}. 
To this end, we suppose that 
the `tunneling' across the horizons is small,
i.e. the $\beta$-Bogoliubov coefficients associated with each horizon are small. 
This is true for $\om/\kappa$ sufficiently large, see \cite{Macher:2009tw}
where it was shown that
$z_\om$ 
and $w_\om$, 
which are related to the Bogoliubov coefficients 
 by $z_\om  = \beta_\om/\alpha_\om$,
behave as $\sim e^{-\pi \om/\kappa}\, (1 - \om / \om_{\rm max})^{1/4}$ where $\om_{\rm max}$ is defined in 
Fig. \ref{figdisp}. 
To proceed we suppose 
that 
the norms  of $z_\om^2$, $w_\om^2$ and $z_\om w_\om$ are much smaller than 1.
In this case, one can expand \eqr{S22} in these three products,
and in 
$\Gamma$, since  the roots 
of \eqr{Upole} are real when $z_\om = w_\om = 0$.


Indeed, 
to zeroth order in these quantities, one gets $\tal_\om \tga_\om= e^{-i \pi}$. 
The norm is trivially constrained by unitarity,
whereas the phase takes its value from the contributions 
stemming from the prefactor of the trapped mode at the two turning points, see \eqr{WKBmodes}.
Taking this into account, $S_{22}= 1$           
gives 
\be
 S^{(1)}_{-\om} - S^{(2)}_{-\om}  + \pi = \int^{R_\om}_{L_\om} dx [- k^{(1)}_{\om}(x) + k^{(2)}_{\om}(x)]+ \pi = 2 \pi\,  n,
\label{BS}
\ee
with  $n \in \mathbb N_1$. 
This is the Bohr-Sommerfeld condition applied 
to the negative frequency mode $\varphi_{-\om}$.
(In fact, subtracting from both $k^{(1)}$ and $k^{(2)}$ the value of $k$ at the turning points, 
$k_\om^{tp}$, the differences $k^{(1)}_\om- k_\om^{tp}$, $k^{(2)}_\om- k_\om^{tp}$ have opposite sign.
Hence, \eqr{BS} contains a sum of two positive contributions, as in the Bohr-Sommerfeld condition.)
We call $\om_a$, $a= 1, 2, ... , N$ the discrete set of frequencies, 
which is finite because no solution exists above $\om_{\rm max}$.

To first order, 
for each $\om_a$, 
one gets a complex phase shift $\delta \la_a = \delta \om_a+ i \Gamma_a$.
 The imaginary shift is 
\ba
 2 \Gamma_a T^b_{\om_a} = \vert z_{\om_a} \vert^2 +  \vert w_{\om_a} \vert^2 +
2 \vert z_{\om_a} w_{\om_a} \vert\,  \cos\psi_a  
= \vert S_{12}(\om_a) \vert^2. 
\label{Gddr}
\ea
The phase in the cosine is 
\be
\psi_a = 
S^{u}_{\om_a} + S^{(1)}_{-\om_a} + \arg\left(z_{\om_a} w_{\om_a}^* \alpha_{\om_a}/\tal_{\om_a}\right),
\ee
and $T^b_{\om_a } > 0$ is the
time for the negative frequency 
partner to make a bounce. It is  given by 
\be
T^b_\om =  \frac{\partial}{\partial{\om}} \left( S^{(2)}_{-\om} - S^{(1)}_{-\om} + \arg \tal_{\om}
\tga_{\om} \right) ,
\label{bouncet}
\ee
evaluated for $\om= \om_a$. The first two terms give the classical (Hamilton-Jacobi) 
time, whereas the last one gives the contribution from the scattering coefficients (which is a small
correction when $ L \Lambda D^{1/2}\gg 1$).\footnote{Using the 
techniques of \cite{Corley,Balbinot:2006ua}, we 
verified that it gives a small correction to the classical actions. We
also obtained a refined expression for \eqr{BS} which goes
 beyond the WKB approximation and which has been validated by the numerical analysis of \cite{Finazzi}.}  
The last equality of \eqr{Gddr} tells us that 
$\Gamma_a$
is linearly related to the norm of the effective $\beta$-Bogoliubov coefficient
of the pair, which obeys $\vert S_{22} \vert^2 = 1+  \vert S_{12} \vert^2 $. 
It is also worth noticing that $S_{12}$ 
fixes the amplitude $\beta_a$ of the leaking mode in \eqr{Upole}.

Because 
$\vert S_{22} \vert^2 = 1+  \vert S_{12} \vert^2 $, 
$\Gamma_a$ defined in \eqr{Gddr} 
is 
positive, thereby implying that $\varphi_a$, the solution of \eqr{Uc}, is a growing mode in time, 
and an ABM in space. 
What distinguishes the present case
from usual resonances characterized by a decay rate ($\Gamma < 0$)
is the fact that the norm of the trapped mode $\varphi_{-\om}^*$
is opposite to that of the leaking wave $\varphi^u_\om$. Even though unitarity in both cases 
implies a decrease of the norm of the trapped mode, in the present case it becomes more negative
whereas in the standard case it tends to zero since it has the same sign as that of $\varphi^u_\om$.
As a corollary of this, the fact that resonances have an opposite sign of $\Gamma$ while satisfying an outgoing 
condition as in \eqr{Uc}
implies that they are not ABM, and therefore not included in the set of modes of \eqr{lindec}. 
This remark applies to fermionic fields and implies that the set of ABM for the fermionic
dispersive field considered in~\cite{Corley:1998rk} and propagating in the BH-WH metric of \eqr{vdparam}, 
is restricted to the continous set of \eqr{lindec}, i.e. positive real frequency modes 
elastically scattered as given in \eqr{RT}.
To complete these remarks, one should notice that $\psi_a$, our decaying modes, 
are also ABM because they obey the incoming condition \eqr{eqfpsi}.

Our treatment 
is similar to the 
interesting analysis presented in \cite{Damour:1976kh}. In that work, the general solution is constructed in terms of WKB waves. Then the Bohr-Sommerfeld 
and the outgoing conditions are separately imposed in the small tunneling approximation, thereby 
fixing both the real part of the frequency $\om_a$ and its imaginary part $\Gamma_a$.
We followed another logic which leads to the same result, namely, 
the requirement that the mode be an ABM gives
\eqr{Uc}, which in turn gives
the complex equation $S_{22}= 1$ that encodes both conditions.
We note that $S_{22}= 1$ does not require that the tunneling amplitudes be small to be well defined.
We also note that our
quantization scheme applies to the cases studied in \cite{Damour:1976kh,Damour:1978ug}
and allows to remove the `formal trick' used 
in the second paper.

\subsection{The density of unstable modes}
\label{density}

To further characterize the instability, 
it is of interest to inquire about the density of unstable modes, 
about the end of the set, its beginning, and about the most unstable mode.

The number of unstable modes will be
either large, or small, depending on the 
value of $L/\om_{\rm max}$. When $L/\om_{\rm max}\gg 1$, the number is large because,
for $\om < \om_{\rm max}$,
the gap between two neighboring modes 
roughly given 
 by $D/L$. 
The end of the set is controlled by $\om_{\rm max}$ as explained
after \eqr{BS}.
It does not
significantly contribute to the instability when $\kappa/\om_{\rm max}\ll 1$,
since $z_\om$ 
and $w_\om$ 
tend to zero as $e^{-\pi \om/\kappa}\, (1 - \om / \om_{\rm max})^{1/4}$ for $\om \to \om_{\rm max}$. 
Thus the 
growth rate $\Gamma_a\to 0$ as $\om \to \om_{\rm max}$. 

The beginning of the set is governed by the first Bohr-Sommerfeld modes.
Their frequency is of the order of $D/L$. 
 A proper evaluation of $\Gamma_a$ 
is harder 
since, generically, 
the coefficients $\alpha_\om$ and $\gamma_\om$ both diverge as
$\sim \om^{-1/2}$. We thus conjecture that the 
most unstable mode is  the first 
mode, or one of the first ones, because the cosine
of \eqr{Gddr} could in some cases lower the value of $\Gamma_1$ below that of the next ones.
We can also characterize the migration of the roots as $L$ increases.
The (total) variation of \eqr{BS} with respect to $L$ and \eqr{bouncet}
tells us that $\partial \om_a /\partial L> 0$. 
From this we conjecture that new unstable modes (obtained by increasing $L$)
appear with $\om_a = 0$. 
This is 
corroborated by the first runs of \cite{Finazzi}.
The determination of $\Gamma_a$ 
for $\om_a \to 0$ is difficult, 
and further study is needed to establish if 
$\Gamma_a $ follows 
the abrupt behavior present in Fig. 1 of \cite{Cardoso:2004nk}.

\subsection{Taking into account the $u-v$ mixing}
\label{uvmix}

A nonvanishing $u-v$ mixing hardly modifies the above results. However the algebra becomes 
more complicated 
because the two nontrivial matrices describing
the propagation across the WH and the BH horizon ($U_1$ and $U_3$) are now the $3\times 3$.
Let us only sketch this enlarged case. 

When considering the in mode $\phi^{u,\, in}_\om$,
in the left region $L$, one has 
$\phi^{u,\, in}_\om = \varphi^u_\om + R_\om \varphi^v_\om$, see \eqr{RT}. 
In the inside region $I$, one has
\be
\phi_\om^{u, \, in} = {\cal A}_\om^u \, \varphi^u_\om +  {\cal A}_\om^v \, \varphi^v_\om +
{\cal B}^{(1)}_\om (\varphi^{(1)}_{-\om})^* + 
{\cal B}^{(2)}_\om (\varphi^{(2)}_{-\om})^* ,
\label{phiuin2}
\ee
and in the external $R$ region, the mode asymptotes to 
$T_\om \, \varphi^u_\om$. 
As before, the coefficients are determined by the matching conditions at the horizons.
Expressing $\phi_\om^{u, \,in}$ in terms of three coefficients associated with the 
three WKB waves, 
$(\varphi^u_\om, \varphi^v_\om, \varphi_{-\om}^*)$,
using the Bogoliubov matrices~\cite{Macher:2009tw} $S_{WH},  S_{BH}$ 
which give, for the WH and the BH respectively,
the coefficients after the scattering given the incident ones,
one has
\ba
\left(\begin{array}{c}  {\cal A}_\om^u \\ R_\om \\ {\cal B}^{(1)}_\om \end{array}\right)
= 
\begin{array}{cc} S_{WH} \end{array}
\left(\begin{array}{c} 1 \\  {\cal A}_\om^v \\ {\cal B}^{(2)}_\om\end{array}\right), 
\label{WH}
\ea
and
\ba
\left(\begin{array}{c}  T_\om \, e^{iS^u_\om}\\  {\cal A}_\om^v \, e^{iS^v_\om}  \\{\cal B}^{(2)}_\om \,
e^{- i S^{(2)}_{-\om}}
\end{array}\right)
= 
\begin{array}{cc} S_{BH} \end{array}
\left(\begin{array}{c}  {\cal A}_\om^u \, e^{iS^u_\om} \\ 0
\\ 
{\cal B}^{(1)}_\om \, e^{-i S^{(1)}_{-\om}}
\end{array}\right). 
\label{BH}
\ea
In the second equation, the exponentials 
arise from the 
propagation from the WH to the 
BH horizon, and from our choice that the phase of  
WKB waves vanishes at the WH horizon.
The six relations fix the six coefficients. $T_\om$ and $R_\om$ characterize the asymptotic 
scattering, see \eqr{RT},
whereas the four others determine $\phi_\om^{u, \, in}$ 
in the inside region. 

To get the complex frequencies $\la_a$, since the weight of the incoming $v$ mode 
is already set to $0$, 
one simply needs to replace the $1$ in 
the right hand side of \eqr{WH} by a $0$, as done in \eqr{Uc}.

\section{Physical predictions}
\label{predic2}

\subsection{Classical settings}
\label{clwp}

From an abstract point of view, any (bounded) initial condition, i.e. the data of the field amplitude 
and its derivative with respect to time, can be translated
into the coefficients $a^i_\om,b_a,c_a$ of \eqr{lindec}. This is guaranteed by the 
completeness of the mode basis.
Yet, it is instructive to study more closely how initial conditions describing a wave
packet initially moving towards the WH-BH pair 
translate into these coefficients. 
This will allow us to relate our mode basis to the wave packet analysis of \cite{Corley:1998rk,Leonhardt:2008js}.
Moreover it is a warming up exercise for the determination of the 
fluxes
in quantum settings. 
For simplicity, we work in the regime in which the number of
unstable modes is large, i.e. $L \om_{\rm max} \gg 1$, and with a wave packet of
mean frequency obeying $\bar \omega \ll \om_{\rm max}$  and $L \bar \om\gg 1$, e.g. 
with $\bar \omega = \sqrt{ \om_{\rm max}/L}$.

We consider a unit norm 
wave packet which is initially in the left region and propagating to the right.
At early times, before it approaches the WH horizon, it can thus be expressed 
in terms of the WKB modes 
of \eqr{WKBmodes} as 
\be
\bar \phi^u_{\bar \om}(t, x) = \int_0^\infty d\om \left(f_{\bar \om, x_0}(\om) 
\, e^{ -i\om t} \varphi^u_\om(x) + c.c. \right).
\label{wp}
\ee
As an example, one can consider the following Fourier components 
\be
f_{\bar \om, x_0}(\om) = \frac{e^{-(\om - \bar \om)^2/2\sigma^2}}{\pi^{1/4} \sigma^{1/2}}\,  e^{i \om t_0}\, 
e^{i S_\om^u(x_0,- L)} .
\ee
The first factor fixes the mean frequency $\bar \om $ and the spread $\sigma$ taken to satisfy $\sigma/\bar \om \ll 1$
and $\sigma L \gg 1$.
The last two exponentials ensure that at $t= t_0$ the incoming wave packet is centered around some $x_0 \ll - L$.
$S^u_\om(x, x')$ is the classical action of the $u$-mode evaluated from $x$ to $x'$, 
see the phase of $\varphi^u_\om$ in \eqr{WKBmodes}.

When decomposing $\bar \phi^u_{\bar \om}$ 
as in \eqr{lindec}, the coefficients are given by the overlaps
\ba
\bar a^u_\om &=& (\varphi_\om^{u, \, in} \vert \bar \phi^u_{\bar \om}) = f_{\bar \om, x_0}(\om), 
\quad \bar a^v_\om =  (\varphi_\om^{v, \, in} \vert \bar \phi^u_{\bar \om}) = 0,  
\nonumber\\
\bar b_a &=& (-i ) \, (\psi_a \vert \bar \phi^u_{\bar \om})  \neq 0 ,
\quad 
\bar c_a = (-i ) \,(\varphi_a \vert \bar \phi^u_{\bar \om}) = 0 . 
\label{overl}
\ea
Hence, at all times, the wave packet can be expressed as
\ba
\bar \phi^u_{\bar \om}(t, x) &=& \int_0^\infty d\om  \left(
f_{\bar \om, x_0}(\om) \, e^{ -i\om t} \phi^{u, \, in}_\om(x) +c.c. \right) 
+
\Sigma_a  \left(\bar  b_a\,  e^{- i \la_a 
t} \varphi_a(x) + c.c. \right). 
\label{wp2}
\ea
The first two coefficients in \eqr{overl} are easily computed and interpreted.
They fix the real frequency contribution of the wave packet.
The last two are very interesting and encode the instability. 
We first notice that, because of \eqr{psivar}, they are given by the 
overlap with the partner wave. Hence the coefficients
$\bar c_a$ identically vanish. Indeed the overlap of our 
wave packet with the 
{growing} modes $\varphi_a$ vanishes since these, by construction,  have no
incoming 
branch. On the contrary $ \bar b_a $, the amplitudes of the growing modes, 
do not vanish since the {\it decaying} modes $\psi_a$ contain an incoming branch, see \eqr{eqfpsi}. 
It is thus through the nondiagonal character of \eqr{psivar} that the instability enters in the game.
In this respect, our analysis differs from Ref.~\cite{Barcelo:2006yi}.
We do not understand the rules adopted in that work. 

To compute 
$\bar b_a$ we use the fact that at time $t_0$ the wave packet 
is localized around $x_0 \ll -L $. Thus we need the behavior of $\psi_a$ in this region. 
Using the results of Sec. \ref{zsmall} and \eqr{WKBmodes} extended to complex frequencies, 
for $\Gamma_a T^b_{\om_a} \ll 1$, 
one gets
\ba
\psi_a(x) &=& \tilde \beta_a\,  \varphi^u_{\la_a}(x)
= \tilde \beta_a\, \varphi^u_{\om_a}(x) \times \exp[{-\Gamma_a  \, t_{\om_a}(x, - L)}],
\label{51}
\ea
where $ t_{\om_a}(x,- L) = \partial_\om S^u(x,-L)> 0$ 
is the 
time taken by a $u$-mode of frequency $\om_a$
to travel from $x$ to $-L$.\footnote{It is interesting to notice 
how $\psi_a$ and $\phi_a$ acquire 
a vanishing norm. Because
the leaking wave has an amplitude $S_{21}\propto \Gamma_a^{1/2}$ 
and decreases in $x$ in $\Gamma_a^{-1}$, its positive contribution to the norm is 
independent of $\Gamma_a$ and cancels out that negative of 
the trapped mode.}
Inserting this result in 
 $(\psi_a \vert \bar \phi^u_{\bar \om})$, using the fact that 
the wave packet has a narrow spread in $x - x_0$ given by $1/(\sigma \partial_\om k^u)$, 
the overlap is approximatly 
\be
\bar b_a = \frac{ \tilde \beta_a^* 
}{\pi^{1/4} (2\sigma)^{1/2}}\, e^{-i(\bar \om - \om_a)t_0}\, e^{ i S^u_{\om_a}(x_0,-L)} \times e^{-\Gamma_a  t_{\om_a}(x_0, - L)},
\ee
The meaning of this result is clear. The spatial decay of $\psi_a$ governed by $\Gamma_a$
has the role to delay the growth of the amplitude of the wave packet until 
it reaches the WH-BH pair. The phase 
ensures the spatial coherence of the propagation. That is, the sum of $a$ in \eqr{wp}
will give constructive interferences along the classical trajectory emerging
from $x_0$ at  
$t_0$.
Let us also mention that, because of \eqr{psivar}, the normalization of $\bar b_a$ is arbitrary,
but the product $\bar b_a \varphi_a$ is well defined and physically 
meaningful.

From this analysis, we see that at early times the wave packet $\bar \phi^u_{\bar \om}$ behaves as described 
in \cite{Corley:1998rk,Leonhardt:2008js}. It propagates freely without any growth until
it reaches the horizon of the WH. Then one piece is reflected and becomes a $v$ mode 
with amplitude $R_{\bar \om}$. The other piece enters in the inside region. 
A later times, a component 
stays trapped and bounces back and forth while its amplitude increases in as $e^{\bar \Gamma t}$.
Every time it bounces 
there is leakage of a $u$ 
wave packet at the 
BH 
horizon, and a $v$ mode at the WH one. 

This back and forth 
semiclassical movement goes on
until the most unstable mode, that with the largest $\Gamma_a$,
progressively dominates and therefore progressively destroys the 
coherence 
of the successive emissions.
In Sec.\ref{density} we saw that the most unstable mode is likely to be 
that characterized by the smallest real frequency, 
$n=1$ in \eqr{BS}. 
This implies that at late time, 
our wave packet will be completely governed
by the corresponding wave $\varphi_1(x)$ (unless of course $\bar b_1 = 0$)
\be
\bar \phi^u_{\bar \om} \to e^{\Gamma_1 (t-t_0)} 
 \times \, {\rm Re}[ e^{-i\om_1 t}\, \bar b_1 \varphi_1(x)]. 
\ee
In the late time limit, its behavior
differs from the above semiclassical one.  
Indeed, in the inside region, one essentially has a standing wave whose amplitude exponentially grows
(it would have been one if the tunneling amplitudes were zero). 
Outside, for $x \gg L$, 
 using $\varphi_1(x) \sim \beta_1 \varphi^u_{\la_1}(x)$ and the same approximation as in \eqr{51},  
one has a modulated oscillatory pattern given by 
\be
\bar \phi^u_{\bar \om} \to e^{\Gamma_1 [(t -t_0) - t_{\om_1}(L,x)]}
\times {\rm Re}[ e^{-i\om_1 t}\, \bar b_1 \, \beta_1 \varphi^u_{\om_1}(x)]. 
\label{aswplt}
\ee 
It 
moves with a speed equal to 
$\sim 1$ in the dispersionless regime, when $\om_1/\Lambda \ll 1$. 
The energy flow is now given by a sine squared of period $2 \pi/\om_1$ 
rather than being composed of localized packets separated by the bounce time $T^b$.

\subsection{Quantum settings}

\subsubsection{The initial state}

Because of the instability, there is no clear definition of what the vacuum state should be.
Indeed, as can be seen from \eqr{Hlindec},
the energy is unbound from below. 
Therefore, to identify 
the physically relevant states, one should inquire what would be the state,
or better the subset of states, which would obtain when the BH-WH pair is
formed at some time $t_0$. If this formation is adiabatic, the initial state
would be close to a vacuum state at that time. 
That is, the expectation values of the square of the various field amplitudes 
would be close to their values in minimal uncertainty states, with no squeezing,  i.e. no anisotropy
in $\phi_\om-\pi_\om$ plane, where $\pi_\om$ is the conjugated momentum of $\phi_\om$. 
Because of the orthogonality of the eigenmodes, 
this adiabatic state would be, and stay, 
a tensor product 
of states associated with each mode separately. 

There is no difficulty to apply these considerations 
to the real (positive) frequency oscillators which are described by 
standard destruction (creation) operators $\hat a_\om^u, \, \hat a_\om^v$ ($\hat a_\om^{u\, \dagger}, 
 \hat a_\om^{v\, \dagger}$). 
The adiabaticity guarantees that one obtains 
a state close to the ground state 
annihilated by the destruction operators. 
Because of the elastic character of \eqr{RT}, for all 
real frequency oscillators, one gets 
stationary vacuum expectation values with no sign of instability.

There is no difficulty either 
for the complex frequency oscillators described by $\hat b_a$ and $\hat c_a$. 
Indeed, as shown in the Appendix, 
one can define two destruction operators $\hat d_{a+}$ and $\hat d_{a-}$, and use them to 
define the state as that annihilated by them 
at $t_0$. In this state, we get the following 
{nonstationary} vacuum expectation values:
\ba
\langle  \hat b_{a'}(t)\,  \hat b^\dagger_a(t') \rangle & =&
 \frac{ \delta_{a', \, a}}{2} \ e^{-i \om_a (t - t')} \, e^{\Gamma_a(t + t' - 2 t_0)}, 
\nonumber \\
\langle  \hat c_{a'}(t)\,  \hat c^\dagger_a(t') \rangle & =&
 \frac{\delta_{a', \, a}}{2} \ e^{-i \om_a (t - t')} \, e^{- \Gamma_a(t + t' - 2 t_0)}  , 
\nonumber \\
\langle  \hat b_{a'}(t)\,  \hat c^\dagger_a(t') \rangle & =& i \,  \frac{ \delta_{a', \, a}}{2} \
 e^{-i \la_a (t - t')}. 
\ea
As expected, the expression in $b b^\dagger$ ($c c^\dagger$) leads to an exponentially growing (decaying) contribution,
whereas the cross term is constant at equal time. 
This behavior is really peculiar to unstable systems.
Even though the metric is stationary, there is no
normalizable state in which the expectation values of the $b,c$ operators are constant.
Stationary states do exist though, but they all have an infinite norm, see the Appendix.

Using \eqr{lindec}, and putting  $t_0 = 0$ for simplicity, 
the two-point function 
in this vacuum state is 
\ba
&\langle \hat \phi(t,x)\,  \hat \phi(t',x') \rangle = \nonumber \\ 
&\int_0^\infty  d\om\, e^{- i \om (t- t')}\, [ \phi_{\om}^u(x) (\phi_{\om}^u(x'))^*  +  \phi_{\om}^v(x) (\phi_{\om}^v(x'))^*  ] \nonumber \\
&\sum_{a=1}^N {\rm Re} \left( e^{- i \la_a (t - t')} \varphi_a(x) \psi^*_a(x') \right. \nonumber \\
& \left. -  e^{- i \la^*_a (t - t')}  \psi_a(x)  \varphi^*_a(x') \right) \nonumber \\
&\sum_{a=1}^N {\rm Re}\left( e^{- i (\la_a t - \la_a^* t')} \varphi_a(x) \varphi^*_a(x') \right.  \nonumber \\
&+ \left. e^{- i (\la^*_a t - \la_a t')} \psi_a(x)  \psi^*_a(x') \right). \\
\label{2ptf}
\ea
We notice that the last term, the second contribution of 
complex frequency modes is real, 
as a classical term (a stochastic noise) 
 would be. We shall return to this point below.
We also notice that the second term is purely imaginary and, when evaluated at the 
same point $x=x'$, it is confined inside the horizons since $\varphi_a$ vanishes for $x\ll -L$
whereas $\psi_a$ does it for $x\gg L$. Thus it will give no asymptotic contribution to local observables.  

\subsubsection
{The asymptotic fluxes}

Our aim is to characterize the asymptotic particle content 
encoded in the growing modes  $\varphi_a$. 
To this end
it is useful to introduce a particle detector localized far away from the BH-WH pair. We take
it to be sitting at $x \gg L$, in the $R$ region on the right of the BH horizon. We assume that 
it oscillates with a constant frequency $\om_0 > 0$, and that 
its coupling to $\phi$ is switched on at
 $t= - \infty$, and switched off suddenly at $t= T \gg t_0$ in order to see how the
response function is affected by the laser effect a finite time after the formation of the BH-WH pair at $t_0= 0$.

When the detector is initially in its ground state,
to second order in the coupling $g$ with 
the field, the probability to find it excited at time $T$ is given by~\cite{Brout:1995rd} 
\ba
P_e(T) &=& g^2 \int_{-\infty}^T dt'  \int_{-\infty}^T dt  \,e^{-i\om_0(t-t')} \,  
 \langle \hat \phi(t,x)\,  \hat \phi(t',x) \rangle ,
\nonumber \\
& = & g^2\,  \Sigma_a \, \vert \beta_a 
\vert^2 \, \,\vert \varphi^u_{\la_a}(x)\vert^2 \, \,
\vert  \int_{t_{\om_a}(L,x)}^T
dt\,  e^{- i(\om_0 - \la_a)t} \vert^2 ,
\nonumber \\
& = & g^2 \, \Sigma_a \, \frac{ \vert \varphi^u_{\om_a}(x)\vert^2  }
{(\om_0 - \om_a)^2 + (\Gamma_a)^2} \,\bar n_{\om_a}(T,x)  , 
\label{rate}
\ea
where 
\be
\bar n_{\om_a}(T,x) = 
\vert \beta_a \vert^2 \, 
e^{2 \Gamma_a (T - t_{\om_a}(L,x))}\
 \vert 1 - \exp  \{ - [\Gamma_a + i (\om_0 - \om_a)](T - t_{\om_a})\} \,  \vert^2 .
\ee

To get the second line of \eqr{rate}, we used $\varphi_a(x) \sim \beta_a \varphi^u_{\la_a}(x)$, 
the fact that the BH-WH pair is formed at $t=0$,
and that it takes a time $t_{\om_a}(L,x)$ for the mode $\varphi^u_{\la_a}$ to reach the detector at $x$. 
To get the third line, we used the inequality $\Gamma_a T^b_{\om_a}\ll 1$ 
as in \eqr{51}. 
%
The meaning of the various factors appearing in 
$P_e$ 
is the following. The sum over $a$ means that all unstable modes
contribute, but the Lorentz functions restrict the significant contributions to frequencies
$\om_a$ near $\om_0$, that of the detector. The prefactor $\vert \varphi^u_{\om_a}(x)\vert^2$ 
depends on the norm of the corresponding mode evaluated at the detector location, as in the usual case. 
The function $\bar n_{\om_a}(T,x)$ acts as 
the number of particles of frequency $\om_a$ received by the detector 
at time $T$, and at a distance $x-L \gg \kappa_B^{-1}$ from the BH horizon. 
It depends on the number initially emitted 
($=\vert \beta_a \vert^2$) multiplied by the exponential governed by $T - t_{\om_a}$, the 
lapse of time since the onset of the BH-WH pair minus the time needed to reach the detector at $x$.

From the 
response function of a localized detector, it is 
clear that  one cannot 
distinguish between the noise due to quanta of the real frequency modes $\phi^u_\om$ and 
that carried by the growing modes $\varphi_a$,  
because both modes asymptote to the WKB waves $\varphi^u_\om$ 
which are asymptotically complete. 
In this respect it is particularly 
interesting to compute the de-excitation probability $P_d$ 
which governs the (spontaneous + induced) decay of the detector. 
It is obtained by replacing $\om_0$ by $- \om_0$ in the first line of \eqr{rate} \cite{Brout:1995rd}.
In that case, one finds that the 
spontaneous decay only comes from the
$\phi^u_\om$ whereas the induced part only comes from the $\varphi_a$.
The induced part equals that of $P_e$ 
since the asymptotic contribution of the $\varphi_a$ to \eqr{2ptf} is real.
 We are not aware of other circumstances 
where orthogonal modes with different eigenfrequency (here $\phi^u_\om$ and $\varphi_a$)
are combined in this way in the spontaneous + induced de-excitation probability $P_d$,
or equivalently, contribute in this way to the commutator and the anticommutator of the field,
i.e. with the $\varphi_a$ only contributing to the latter. (For damped modes, 
the commutator and the anticommutator are related differently, see e.g. Appendices A and B 
in \cite{Parentani:2007uq}.) The lesson we can draw is the following: even though the modes $\phi^u_\om$ are orthogonal to the growing 
modes $\varphi_a$, their respective contribution to the asymptotic particle content 
cannot be distinguished by external devices coupled to the field.

It is also interesting to compute
the  asymptotic outgoing energy flux $ \langle \hat T_{uu}(t,x)  \rangle$, 
where $T_{uu} = [(\partial_t - \partial_x)\phi]^2$. 
At large distances in the $R$ region, using $\varphi_a \sim \beta_a \varphi^u_{\la_a}$
and \eqr{2ptf},
the renormalized value of the flux is
\ba
 \langle \hat T_{uu}(t,x)  \rangle &=& 
\Sigma_a 
\vert (\partial_t - \partial_x)\,  e^{- i \la_a t}\varphi_a(x)\vert^2 
\nonumber\\
&= & \Sigma_a 
\, \vert \beta_a \vert^2  \, 
\vert (\partial_t - \partial_x) \, 
e^{- i \la_a t}\varphi^u_{\la_a}(x)\vert^2 .
\label{Tuu}
\ea
It only depends on the discrete
set of complex frequency modes.
Yet, because of the imaginary part of 
$\la_a$ 
defines a width 
$= \Gamma_a$,  the spectrum of real frequencies $\om$ is continuous,
as can be see in \eqr{rate}.
In fact 
the observables can either be written in terms of a discrete sum
over complex frequencies, or 
as a continuous integral of a sum of Lorentz functions
centered at $\om_a$ and of width $\Gamma_a$. 
However this second writing 
is only approximative and requires that the inequality $\Gamma_a T^b_a \ll 1$ 
of Sec. \ref{zsmall}
be satisfied to provide a reliable approximation. 

\subsubsection{The correlation pattern}

As noticed in \cite{Corley:1998rk}, because of the bounces of the trapped modes,
the asymptotic fluxes possess non-trivial correlations on the {\it same} side of the horizon,
and not across the horizon as in the case of Hawking radiation without 
dispersion~\cite{Massar:1996tx,Brout:1995rd,Balbinot:2007de},
or in a dispersive medium~\cite{Brout:1995wp}.
These new correlations 
are easily described in the wave packet 
language of that reference, or that of Sec. \ref{clwp}.
%
When using 
frequency eigenmodes, 
they can be recovered through constructive interferences, 
as in Sec.IV F. of \cite{Macher:2009nz}.
Indeed, when the complex frequencies modes
form a dense set so that the dispersion of the waves 
can be neglected, 
the sum over $a$ in 
\eqr{2ptf}  
constructively interferes at equal time for two different positions $x$ and $x'$ separated by a propagation time 
\ba
\partial_\om S^u(x,x') = \int_x^{x'}\! dx\,  \partial_\om k_\om^u(x) 
= T^b_\om,
\ea
where  $T^b_\om$ is the bounce time of \eqr{bouncet}.
This is because the differences $\om_a - \om_{a+n}$ are 
 equal $n\times 2\pi/T^b_\om$ 
since the $\om_a$ 
are solutions of 
\eqr{BS}. 
With more precision, the conditions for having these multimodes 
interferences
are, on one hand, $\om_{\rm max}/\kappa \gg 1$ so that dispersion hardly affects the modes
and, on the other hand, $\kappa L \gg 1$, so as to have many modes for $\om$ 
below the Hawking temperature $ \sim \kappa$. (For higher frequencies $\vert \beta_a \vert^2$, 
which governs the intensity of the correlations, is exponentially damped.) 

However, since dispersive effects grow  
and since 
the most unstable mode progressively 
dominates 
the two-point function of \eqr{2ptf}, 
at sufficiently large time the above multimode coherence will be destroyed
and replaced by the single mode coherence of the most unstable one. 
This is unlike what is obtained when dealing with a single BH or WH horizon because in that case~\cite{Massar:1996tx}
the pattern is stationary, and all frequencies 
steadily contribute (significantly for $\om \leq \kappa$).
In the present case, at late time, if $\phi_1$ is the most unstable one,
the correlation pattern is given by 
\be
\langle \hat \phi(t,x)\,  \hat \phi(t',x') \rangle = e^{\Gamma_1 (t + t')}\times  {\rm Re}\left( e^{- i \om_1 (t - t')}\, 
 \varphi_1(x)  
\varphi^*_1(x')
 \right). 
\label{2ptfbis}
\ee
The asymptotic pattern, for $x \gg L$, 
is obtained using $ \varphi_1(x) \sim \beta_1 e^{- \Gamma_1 t_{\om_1}(L,x)} \varphi^u_{\om_1}(x) $.
It is very similar to that of \eqr{aswplt} found by sending 
a classical wave packet, 
see Appendice C of \cite{Macher:2009nz} for a discussion of the correspondence 
between statistical correlations encoded in the two-point function and
deterministic correlations encoded in the mean value when dealing with a wave packet described by 
a coherent state.  

So far we worked under the assumption that the $u-v$ mixing coefficients are negligible.
When taking them into account, one obtains a richer pattern which is 
determined by the complex frequency modes solutions of Sec. \ref{uvmix}.

\subsubsection{The small supersonic region limit}

When $L$ of \eqr{vdparam} (or $D= \vert v_+\vert /c - 1$)
decreases, the number of solutions of \eqr{BS} diminishes.
Therefore, in the narrow supersonic limit $ \kappa L \to 0$,
there is a threshold value for $\kappa L$ given $D$, below which
there is no solution. 
In that case, there are no unstable mode, and
no laser effect. In fact no flux is emitted, and this even though the
surface gravity of the BH (and that of the WH) is not zero. The reason is that 
there is no room for the negative frequency modes $\phi_{-\om}$ to exist. 
In agreement with the absence of radiation, 
the entanglement entropy of the BH~\cite{Bombelli:1986rw} would vanish, because it accounts 
for the number of entangled modes across the horizon and thus of opposite frequency, see 
\cite{Jacobson:2007jx} for the effects of dispersion on the entanglement entropy. 

\subsubsection{Comparison with former works}


It is instructive to compare our expressions to those obtained in \cite{Corley:1998rk} and in \cite{Leonhardt:2008js}. 
Our expressions differ from theirs 
because the discrete character of the set of 
complex frequency modes
was ignored in these works. 
%
As a result, a continuous spectrum was obtained. Yet this spectrum 
possesses rapid superimposed oscillations stemming from the interferences that are
present in $S_{21}(\om)$ of \eqr{S22}. 
{\it A priori} one might think that they could coincide with our
frequencies $\om_a = {\rm Re }\la_a$.
However, as noticed in \cite{Leonhardt:2008js}, their value 
are insensitive the phase governed by 
$S^{(2)}_\om$, whereas it plays a crucial role in \eqr{BS}. 
We found no regime in which the two sets could approximatly agree.
Therefore, as far as the fine properties of the spectrum are concerned, 
the predictions of \cite{Corley:1998rk,Leonhardt:2008js} are not trustworthy.


Nevertheless, when the density of complex frequency modes is high, 
and when ignoring these fine properties, 
the average properties derived using
\cite{Corley:1998rk,Leonhardt:2008js} coincide with ours. Indeed when considering the mean flux 
in frequency intervals $\Delta \om \gg 1/L$, the rapid oscillations found in \cite{Corley:1998rk,Leonhardt:2008js}
are averaged out. As a result the mean agrees with that over the contributions 
of complex frequency modes. This can be explicitly verified by comparing the 
norm of our discrete modes 
$\varphi_a, \psi_a$ with the continuous norm of the negative frequency modes used in \cite{Corley:1998rk,Leonhardt:2008js}. 
In the limit of Sec. \ref{zsmall}
the relevant contribution to the overlap $(\psi_a\vert \varphi_a)$ 
comes from the negative frequency mode $\varphi_{-\om}$. 
Moreover $\varphi_a$ and $\psi_a$ are given by a sum of (normalized) WKB waves 
$\varphi^{(1)}_{-\om_a}$ and $\varphi^{(2)}_{-\om_a}$ 
times a prefactor $= \sqrt{2\pi/T^b_{\om_a}}$ where $T^b_{\om_a}$ is the bounce time
given in \eqr{bouncet}. This is just what is needed for approximating
the discrete sum in \eqr{2ptf} 
by a continuous integral $\int \! d\om$ with a measure equal to 1. 
 
In conclusion, when 
the number of bound modes becomes small, 
the difference between our 
description and the continuous one 
increases. 
This difference is maximal 
 when there are no solution of \eqr{BS}. In this case, 
no radiation is emitted, something which 
cannot be derived by the continuous approach of \cite{Corley:1998rk,Leonhardt:2008js}. 

\section{Conditions for having a laser effect}
\label{condi}

Having understood the black hole laser effect, 
it is worth identify in more general terms 
the conditions under which a laser effect would develop.
We define a `laser effect' by the fact that a free field possesses complex frequency ABM
while being governed by a quadratic hermitian Hamiltonian, as in \eqr{Hphi}, and a conserved scalar product, 
as in \eqr{KGnorm}. The field thus obeys
an equation which is stationary, homogeneous, and second order in time. 
Let us note that this type of instability is 
often referred to as a {\it dynamical instability}~\cite{Gardiner,Barcelo:2006yi,Richartz:2009mi}, 
a denomination which indicates 
that quantum mechanics is not needed to describe/obtain it. Let us also note that we do not consider the 
case where the frequency of the ABM is purely imaginary. 
Such dynamical instability seems to belong to another class than that we are considering,
see the end of this subsection for more discussion on this.

Using semiclassical concepts, the conditions for obtaining complex frequency ABM are the following :
\begin{itemize}
\item For a finite
range of the real part of frequency $\om$,  
WKB solutions with both signs of norm should exist, 
or equivalently, positive norm WKB solutions should exist for both sign of $\om$. 
This is a rather strong condition which requires that the external field (gravitational or electric) 
must be strong enough for this level crossing to take place, i.e. 
for the general solution be a superposition as in \eqr{phiuin}.

\item These WKB solutions of opposite norm must mix when considering the exact solutions of the mode equation. 
In other words there should be connected by a nonzero tunneling amplitude. This is a very weak condition
as different WKB branches are generally connected to each other.
In our case it means that $z_\om$ and $w_\om$ appearing in \eqr{S22} should not vanish. 

\item One of these WKB solutions must be trapped so that the associated 
wave packets will bounce back and forth.
 This is also a rather strong condition.

\item The depth of the potential trapping these modes should be deep enough so that at least
one pair of bound modes can exist, see \eqr{BS}. This condition is rather mild once the first three are satisfied. 

\end{itemize}
When these conditions are met for a sufficiently wide domain of frequency $\om$, 
they are sufficient to get a laser effect, and
they apply both when the external region is finite~\cite{Gardiner} or infinite.
Being based on semiclassical concepts, {\it stricto sensu}, they cannot be considered as necessary.
But we are not here after mathematical rigor, rather we wish to identify
the relevant conditions in physically interesting situations.

In this respect, it should 
be noticed that when only the first two conditions are satisfied, 
one obtains a {\it vacuum instability}~\cite{Brout:1995rd}, also called a {\it superradiance} in the
context of rotating bodies~\cite{Bekenstein:1998nt,Richartz:2009mi}. 
Hence, whenever there is a vacuum instability, one can engender a laser effect by 
introducing a reflecting condition, as was done in \cite{Kang:1997uw,Cardoso:2004nk}, or by
modifying the potential, so that the last two conditions are also satisfied. It should be clear that 
when the laser effect takes place, it replaces the vacuum instability rather than occurs together with it. 
Indeed, as proven in Sec. \ref{zsmall}, the frequency of the trapped modes
are generically complex. The possibility of having a trapped mode
(subjected to a vacuum instability prior introducing the reflecting condition) 
with a real frequency is of measure zero,
as two conditions must be simultaneously satisfied.
To give an example of the replacement of a vacuum instability by a laser effect,
let us consider the archetypal case 
of pair production in a static electric field studied by Heisenberg~\cite{Heis} and Schwinger~\cite{Sch}. 
In that case, one obtains a laser effect 
by replacing the Coulomb potential $A_0 = Ex$ by
$A_0 = E \vert x \vert$ which traps 
 particles of charge $q$ for $qE < 0$, for frequencies $\om < - m$ where $m$ is their mass. 
We hope to return to such pedagogical examples in the near future.

In conclusion, we make several remarks.
Even though the complex frequency ABM are orthogonal
to the real frequency modes, as is it guaranteed by \eqr{ortho}, 
the {\it asymptotic} quanta associated with these 
modes are not of a new type but are, as we saw, superpositions of 
the standard ones associated with real frequency modes.
If laser instabilities can be studied in classical terms, the quantum aspects are 
not washed out. 
For instance, when considering a charged field, the charge received as infinity is
still quantized, albeit its mean value is described by a complex frequency ABM.
Moreover, in all cases, when the instability ceases,
the number of emitted quanta is a well-defined observable governed by standard destruction/creation operators 
as those appearing in \eqr{HCa}.
From this it
appears that dynamical instabilities governed by an ABM with a purely imaginary 
frequency, as e.g. the Gregory-Laflamme instability~\cite{Gregory:1993vy}, 
belong to another class since this asymptotic decomposition in terms of quanta 
 does not seem available. Whether it could nevertheless
makes sense to quantize such instability is a moot point. 

\section{Conclusions} 

We showed that the black hole laser effect should be 
described in terms of a finite and discrete set of complex frequency modes which asymptotically vanish.
We also showed that these modes are orthogonal to the continuous set of
 real frequency modes 
which are only elastically scattered, 
and which therefore play no role in the laser effect.  
In Sec. \ref{modeprop}, using the simplifying assumption that the near horizon regions are thin,
we determined the set of complex frequencies and the properties of the modes using the 
approach that combines and generalizes  \cite{Leonhardt:2008js} and \cite{Damour:1976kh}. 

We described 
how an initial wave packet is 
amplified 
as it propagates in the BH-WH geometry. 
When the density of complex frequency modes is high we recovered the
picture of \cite{Corley:1998rk} at early times. Instead, at late time, or when the density is low, 
the successive emissions of distinct wave packets associated with the bouncing trajectories 
are replaced by an oscillating flux governed by the most unstable mode.

We then computed in quantum settings how the
growth of the complex frequency modes determine the asymptotic 
fluxes when the initial state at the formation of the BH-WH pair is vacuum. 
Because of the width associated with the instability, the spectral properties of
the fluxes are continuous albeit they arise from a discrete set of modes.
The properties we obtained significantly differ from those found in \cite{Corley:1998rk,Leonhardt:2008js}.
%
We also briefly described the properties of the correlation pattern at early times when the number of 
complex frequency modes is large, and at late time when only the most unstable mode contributes. 
When the supersonic region between the two horizons is too small so that 
there is no solution to \eqr{BS}, 
we concluded that there is no instability,
that no flux is emitted, and that the entanglement entropy vanishes.
Finally, in Sec.~\ref{condi} we gave the general WKB conditions under which a laser effect would
obtain starting from the standard concepts that govern a vacuum instability in Quantum Field Theory.

This work poses several 
questions which deserve further study.
In Bose condensates, the backreaction due to the instability and the suppression of the instability
 could be computed using the Gross-Pitaevskii equation. 
Fermionic fields~\cite{Corley:1998rk,Damour:1978ug} should be 
further studied to reveal the roles of the $N$ pairs of complex frequency modes, which are not ABM,
but which indicate that the naive vacuum will decay in the lowest energy state vacuum plus $N$ asymptotic quanta
by spontaneously emptying the $N$ Dirac holes which are trapped inside the horizons.

\begin{acknowledgments}
R.P. would like to thank N. Deruelle, T. Jacobson, C. Mayoral, and J. Mourad for stimulating discussions
over the last year. We also wish to thank R. Balbinot, C. Barcelo, I. Carusotto, A. Fabbri, S. Liberati, and A. Recati for remarks in the last stages of this work. We are grateful to I. Carusotto
for critical remarks on a early draft of the present work, and to S. Finazzi for sharing the numerical results
of \cite{Finazzi}. 

\end{acknowledgments}

\appendix
\section{Upside down harmonic oscillators\label{upsidedho}}


\subsubsection{Real upside down oscillator}

We review the quantization of upside down 
oscillators.
To begin with, we start with a single real upside down harmonic oscillator.
Its Hamiltonian is 
\be
H = \frac{1}{2} (p^2 - \Gamma^2 q^2),
\label{Hho}
\ee
when written in terms of position $q$ and conjugated momentum $p$,
obeying the standard equal time commutator (ETC) $[q, p] = i$. Introducing the 'null' 
combinations 
\be
b = \frac{1}{\sqrt{2 \Gamma}} (p + \Gamma q), \quad c = \frac{1}{\sqrt{2 \Gamma}} (p - \Gamma q),
\label{realbc}
\ee
one gets 
\be
H = \frac{\Gamma}{2}( b c + c b). 
\label{Hho2}
\ee
One verifies that they obey the ETC $[b,c] = i$. The ordering of $b$ and $c$ in $H$
follows from that of \eqr{Hho}. The equations of motions are
\be
\dot b = (-i) [b, H] = \Gamma\,  b , \quad \dot c = (-i) [c, H] = - \Gamma \,  c, 
\ee 
thereby establishing that $b$ ($c$) is the growing (decaying) mode 
$b= b_0 \, e^{\Gamma t}$, ($c= c_0\,  e^{-\Gamma t}$). 

It is now relevant to look for stationary states. In the $b$-representation ($c = -i\partial_b$), 
the stationary Schroedinger equation $H \Psi_E = E \Psi_E$ reads
\be
- i b\partial_b \Psi_E(b) = (\frac{E}{\Gamma}- \frac{i}{2}) \Psi_E(b).
\ee
Solutions exist for all real values of $E$, and the general solution is 
\be
\Psi_E(b) = A_E \, \theta(b) (b)^{i E/\Gamma - 1/2} + B_E \, \theta(-b) (-b)^{i E/\Gamma - 1/2}.
\ee
Since the spectrum is continuum, one should adopt a Dirac delta normalization 
$\langle \Psi_{E'} \vert \Psi_E \rangle = \delta(E - E')$. This gives 
\be
\vert A_E \vert^2 + \vert B_E \vert^2 = \frac{1}{2 \pi \Gamma}. 
\ee
Imposing that the solution be even in $q$ ($p$, $b$, or $c$) imposes $A_E = B_E$. 
The important lesson one should retain is that there is no square integrable 
stationary states. Therefore, in all physically acceptable states (i.e. square integrable) 
the expectation values of $q^2 + p^2$
will exponentially grow $\sim e^{2 \Gamma t}$
at late time. 

\subsubsection{Complex oscillators}

More relevant for us is the complex upside down harmonic oscillator. It can be described by the
complex variables $q= q_1 + i q_2$, $p= p_1 + i p_2$ where $q_1,p_1$ and $q_2,p_2$
are hermitian and obey the 
standard ETC given above. We then introduce the complex $b$ and $c$ variables
\be
b = \frac{1}{\sqrt{4 \Gamma}} (p + \Gamma q), \quad c = \frac{1}{\sqrt{4 \Gamma}} (p - \Gamma q),
\ee
which are normalized so that they obey the ETC
\be
[b,c^\dagger] = i.
\label{ETC2}
\ee

We then look for the (hermitian) Hamiltonian which gives the following equations 
\be
\dot b = (-i)[b,H] = - i \la b, \quad \dot c = (-i)[c,H] = - i \la^* c,
\label{bctev}
\ee
where $\la = \omega + i \Gamma $. 
It is given by
\ba
H &=& - i\la \, c^\dagger b + i \la^* \, b^\dagger c .
\label{HC}
\ea
It is instructive to reexpress this system is term of a couple of destruction and creation operators
$d_-,d_-^\dagger$, $d_+,d_+^\dagger$ which are given, at a given time $t_0$, by
\be
b = \frac{1}{\sqrt{2}}(d_+ - i d_-^\dagger), \quad 
c = \frac{1}{\sqrt{2}}(- i d_+ + d_-^\dagger).
\ee
They obey the standard commutation relations $[d_i,d_j^\dagger] = \delta_{ij}$, 
and \eqr{HC} reads
\ba
H &=& \om \, (d_+^\dagger d_+ - d_-^\dagger d_-) + \Gamma \, ( d_- d_+ + d_+^\dagger d_-^\dagger),
\nonumber\\
&=& H_0 + H_{sq} .
\label{HCa}
\ea
In the first term one recovers the standard form of an Hamiltonian is the presence of stationary modes
with opposite frequencies~\cite{Macher:2009nz}. The second term 
induces a squeezing of the state of the $d_-, d_+$ oscillators which grows linearly with time. 

To set initial conditions, and to be able to read the result of the instability in terms of quanta, 
it is appropriate to use this decomposition of $H$ and to work in the `interacting' picture
where the operators $b,c$ 
only evolve according to $H_0$, and where 
the squeezing operator acts on the state of the field. Indeed, in this picture the
states can be expressed at any time as a superposition of states with a definite occupation numbers $n_-$ and $n_+$.

\subsubsection{The hermitian conjugated $S^\dagger$}

To make contact with the treatment of Sec. \ref{Sma}, 
it is appropriate to express the time evolution of linear operators 
in a $S$-matrix language. We introduce the operator
\be
A[Z(t)] = x(t) \, b_0 + y(t) \, c_0 = \begin{pmatrix} b_0 & c_0 \end{pmatrix}\begin{pmatrix} x(t) \\ y(t) \end{pmatrix},
\ee
where $b_0$ and $c_0$ are the operators $b,c$ evaluated at $t_0$.
Using \eqr{bctev}, the time evolution of 
 $Z = (x, y)$ is 
\be
i \partial_t \begin{pmatrix} x \\ y \end{pmatrix} = H \begin{pmatrix} x \\ y\end{pmatrix}
\ee
where the $2\times 2$ matrix $H$ is 
${\rm diag}(\lambda, \lambda^*)$. 
By definition, the $S$ matrix 
 brings 
$Z$
from $t_0$ to $t_0 + t$. 
It is given by
\be
S = \begin{pmatrix} e^{-i\lambda t} & 0 \\ 0 & e^{-i\lambda^* t} \end{pmatrix}.
\label{Sm}
\ee
To define $S^\dagger$ we need to refer to the matrix encoding the ETC of $b$ and $c$, see \eqr{ETC2}:
\be
Q = \begin{pmatrix} [b,b^\dagger] &  [b,c^\dagger] \\   [c,b^\dagger]&  [c,c^\dagger] \end{pmatrix} = 
\begin{pmatrix} 0 & i  \\ - i &  0 \end{pmatrix}.
\ee
This matrix defines a scalar product in the space of the vectors $Z$ through
\be
[A(Z'), \, A(Z)^\dagger] = (Z \vert Z') =  \begin{pmatrix} x^* & y^* \end{pmatrix}  Q  
\begin{pmatrix} x' \\ y' \end{pmatrix} .
\ee
The hermitian conjugate defined by $(S Z \vert Z') = (Z \vert S^\dagger Z')$ 
obeys
\be
Q\, S^\dagger = S^{*\, T} \, Q. 
\ee
In the present case, using \eqr{Sm},
we find
\be
S^{\dagger}(\lambda) 
= \begin{pmatrix}  e^{i\lambda t} & 0 \\ 0 & e^{i\lambda^* t} \end{pmatrix} = \left[S(\lambda^*)\right]^*.
\label{Sdag}\ee
In the 
case where $S$ would have been nondiagonal, we should also transpose the matrix:
\be
S^{\dagger}(\lambda) = \left[S(\lambda^*)\right]^{*T}.
\label{Sdag2}
\ee
We see that the expression of the components of $S^\dagger$ in terms of those of $S$
is unusual because the norm of $Z$, instead of being $\vert x\vert^2 + \vert y \vert^2$,
it is of the form  Im$(x y^*)$.
However, when $\lambda$ is real, \eqr{Sdag} gives the usual relation. 
Hence, for complex $\lambda$, 
it can be viewed as an analytical continuation of the real frequency case. 

The hermicity of the Hamiltonian guarantees that the $S$ matrix is unitary in the sense 
that the scalar product $(Z' \vert Z)$, or the  ETC relations \eqr{ETC2}, are conserved.
One verifies that \eqr{Sdag} implies the usual unitarity condition
\be
S^{\dagger}(\lambda) \, S(\lambda) = I,
\ee
for $\lambda$ real and complex. 

The above analysis applies to the modes $\varphi_a, \, \psi_a$
when considering the field operator restricted to the $a$ sector, 
i.e. $A(Z_a) = x \, b_a \varphi_a + y \, c_a \psi_a$,
because $\varphi_a, \, \psi_a$ have complex conjugated frequencies and because $b_a, c_a$ obey 
\eqr{combc}. Hence, 
the hermitian conjugated 
matrix appearing in 
\eqr{eqfpsi}
should be understood
as in \eqr{Sdag2}.

\end{document}